\documentclass[%
reprint,
superscriptaddress,
amsmath,amssymb,
aps,
pra,
longbibliography,
floatfix,
]{revtex4-1}

\usepackage[T1]{fontenc}
\usepackage{upgreek}
\usepackage{graphicx}
\usepackage{dcolumn}
\usepackage{bm}
\usepackage{braket} 
\usepackage[%
  colorlinks=true,
  urlcolor=blue,
  linkcolor=blue,
  citecolor=blue
]{hyperref}

\begin{document}

\preprint{APS/123-QED}

\title{Microscale Fiber-Integrated Vector Magnetometer with On-Tip Field Biasing using NV Ensembles in Diamond Microcystals}

\author{Jonas Homrighausen}
\affiliation{Department of Physics Engineering, FH M\"unster - University of Applied Sciences, Stegerwaldstrasse 39, D-48565 Steinfurt, Germany}
\author{Frederik Hoffmann}
\affiliation{Department of Electrical Engineering and Computer Science, FH M\"unster-University of Applied Sciences, Stegerwaldstrasse 39, D-48565 Steinfurt, Germany}%
\author{Jens Pogorzelski}
\affiliation{Department of Electrical Engineering and Computer Science, FH M\"unster-University of Applied Sciences, Stegerwaldstrasse 39, D-48565 Steinfurt, Germany}
\author{Peter Gl\"osek\"otter}
\affiliation{Department of Electrical Engineering and Computer Science, FH M\"unster-University of Applied Sciences, Stegerwaldstrasse 39, D-48565 Steinfurt, Germany}%
\author{Markus Gregor}
\email{markus.gregor@fh-muenster.de}
\affiliation{Department of Physics Engineering, FH M\"unster - University of Applied Sciences, Stegerwaldstrasse 39, D-48565 Steinfurt, Germany}%

\date{\today}

\begin{abstract}
In quantum sensing of magnetic fields, ensembles of nitrogen-vacancy centers in diamond offer high sensitivity, high bandwidth and outstanding spatial resolution while operating in harsh environments. Moreover, the orientation of defect centers along four crystal axes forms an intrinsic coordinate system, enabling vector magnetometry within a single diamond crystal. 
While most vector magnetometers rely on a known bias magnetic field for full recovery of three-dimensional field information, employing external 3D Helmholtz coils or permanent magnets results in bulky, laboratory-bound setups, impeding miniaturization of the device. 
Here, a novel approach is presented that utilizes a fiber-integrated microscale coil at the fiber tip to generate a localized uniaxial magnetic field. The same fiber-tip coil is used in parallel for spin control by combining DC and microwave signals in a bias tee. 
To implement vector magnetometry using a uniaxial bias field, the orientation of the diamond crystal is pre-selected and then fully characterized by rotating a static magnetic field in three planes of rotation. 
The measurement of vector magnetic fields in the full solid angle is demonstrated with a shot-noise limited sensitivity of $19.4\:\textrm{nT/Hz}^{1/2}$ and microscale spatial resolution while achieving a fiber sensor head cross section of less than $1\textrm{mm}^2$.
\end{abstract}

\maketitle


\section{\label{sec:intro}Introduction}

In recent years, the negatively charged nitrogen-vacancy center (NV center) in diamond has been established in the field of quantum sensing, finding its way from laboratory to field-tested applications. Amongst other promising quantum magnetometer candidates, e.g. SQUIDs and alkali vapour cells, 
NV centers are particularly attractive in situations, where high spatial resolution ranging down to atom size \cite{tetienne2014nanoscale,kucsko2013nanometre,wang2021nanoscale}, room temperature operation or high bandwidth \cite{schloss2018simultaneous,wang2021nanoscale,shao2016wide} is required, while surpassing the sensitivity of classical magnetometers like Hall sensors. Sensitivities in the order of pT/Hz$^{1/2}$ have been demonstrated \cite{graham2023fiber,wolf2015subpicotesla,barry2016optical}, down to the femtotesla regime with the aid of magnetic flux concentrators \cite{xie2021hybrid, fescenko2020diamond}.
Furthermore, the solid-state material platform offers a potential high degree of integration and miniaturization of the sensor device \cite{kim2019cmos,pogorzelski2024compact}, as well as operation under extreme conditions like high temperature \cite{toyli2012measurement,barson2019temperature,plakhotnik2010luminescence} and high pressure \cite{doherty2014electronic,barson2017nanomechanical}. 
Numerous of recent experiments were performed in the laboratory using bulky and cost-intensive setups. Consequently, efforts have been made to integrate the setup into a portable in miniaturized device such as a fiber tip sensor \cite{duan2019efficient,homrighausen2023edge,filipkowski2022magnetically,zhang2021robust,dix2024miniaturized}, which allows versatility and a wide range of applications due to the spatial separation of sensor material and optical and electronic components.

\begin{figure}[ht]
\centering
  \includegraphics[width=0.5\textwidth]{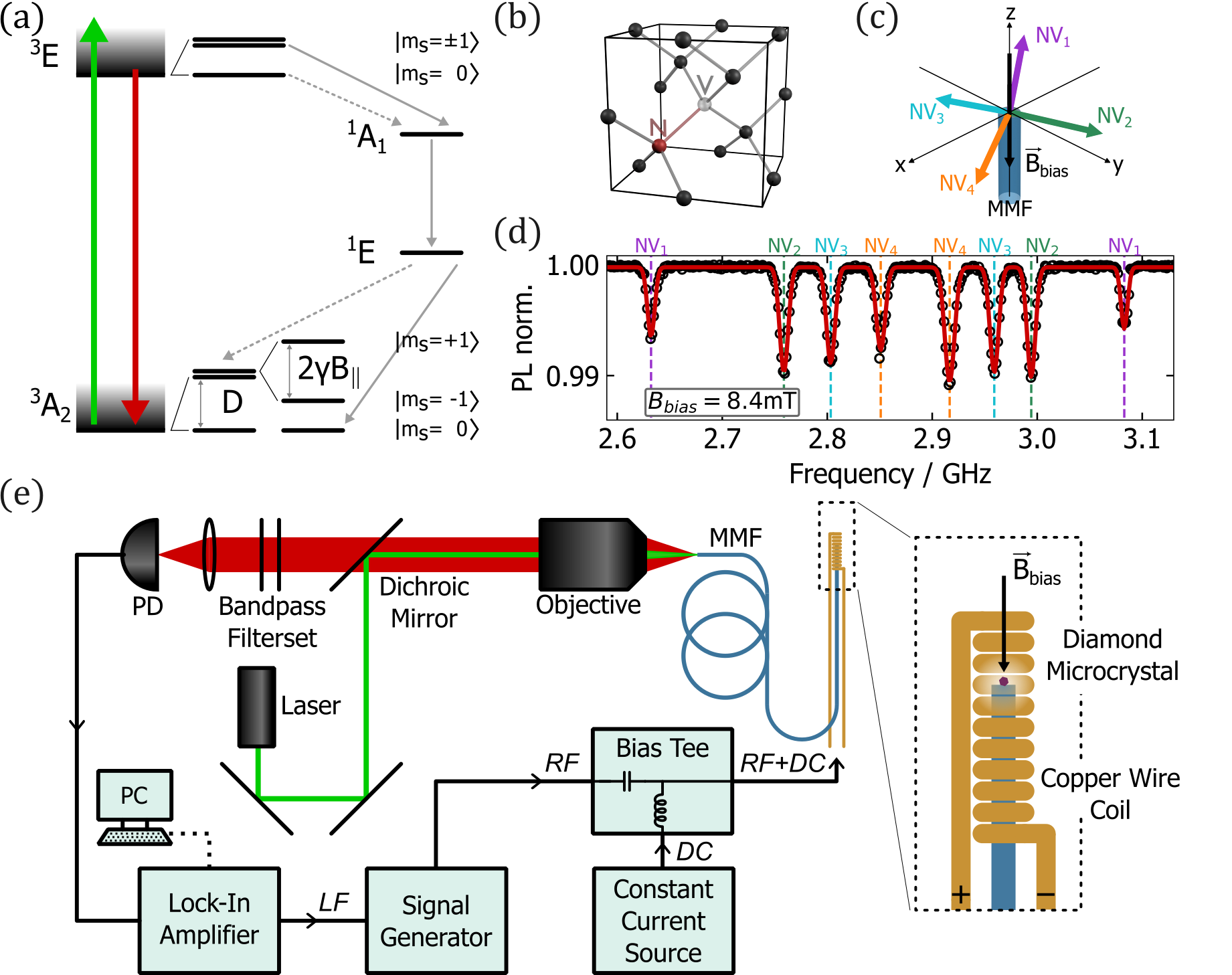}
  \caption{(a) Energy level diagram of the negatively charged nitrogen-vacancy (NV) center in diamond. The spin dependent fluorescence will decrease when the ground state electron spins are driven in resonance to the $\ket{m_S=0} \rightarrow \ket{m_S = \pm 1}$ transitions with $D=2.87\:\textrm{GHz}$ in zero field. The degeneracy of the $\ket{m_S=\pm 1}$ spin states is lifted with a magnetic field parallel to the NV symmetry axes $B_\parallel$. (b) NV center in the diamond crystal lattice, with the NV symmetry axes emphasized in red. Since the nitrogen atom N can occupy each of the four lattice sites surrounding the vacancy V, all four NV orientations will be equally present in NV ensembles. (c) NV ensemble on the tip of a fiber aligned in the $z-$axis. The NV axes NV$_{i}$ form an inherent tetrahedral coordinate system. (d) Optically Detected Magnetic Resonance of the NV ensemble shown in c. The frequency shifts of the resonances corresponding to NV$_{i}$ strongly depend on their angle to the magnetic field $B_\mathrm{bias}.$ (e) Experimental setup with fiber-integrated microscale magnetic coil. The NV diamond is attached to the fiber facet of a multimode optical fiber. The magnetic coil is simultaneously used for active field biasing and spin control using a bias tee combining the signals.}
  \label{fig:fig1}
\end{figure}

The NV center is an optically active crystal defect in the diamond lattice that consists of a substitutional nitrogen atom and an adjacent vacancy. Its magnetic sensing capability is given by the interaction of a magnetic field with the electron spin known as the Zeeman effect. 
This shift of the electron spin sublevels in the NV ground state can be read out in optically detected magnetic resonance (ODMR) experiments. Here, manipulation of the electron spin with microwave (MW) frequencies resonant to the $\ket{m_S=0}\rightarrow \ket{m_S=\pm1}$ electron spin transitions in the ground state $^3A_2$ will decrease the photoluminescence intensity (PL) emitted by the NV centre by increased non-radiative decay via the $^1A_1$ singlet state (see Fig.\:\ref{fig:fig1}(a)). 

The ground state of the NV center exhibits a zero field splitting of $D=2.87\:\textrm{GHz}$ between the $\ket{m_S=0}$ and $\ket{m_S=\pm 1}$ electron spin states. When subjected to a (low) magnetic field, the $\ket{m_S=\pm1}$ spin states undergo a splitting denoted by $\Delta f \approx 2\sqrt{(\gamma B_\parallel)^2+E^2}$, where $B_\parallel = B\cos{\vartheta}$ is the field component along the NV symmetry axis (see Fig.\:\ref{fig:fig1}(b)), $\gamma= 28\:\textrm{MHz/mT}$ is the gyromagnetic ratio and $E$ is a strain-dependent parameter that can lift the degeneracy of the $\ket{m_S=\pm 1}$ electron spin states in zero field conditions due to local crystal strain. 
The dependence of the shift on the axial field component $B_\parallel$ introduces the directionality that enables vector magnetometry. Note that, as discussed in Appendix \ref{sec:appendix_transf}, in higher magnetic fields, the influence of the non-axial component $B_\perp$ of the magnetic field on the transitions frequencies becomes non-negligible.

Due to the $C_{3v}$ symmetry of the diamond crystal lattice, the orientation of a single NV center can align along one of four possible axes that correspond to the Miller indices $[111],[\bar1 \bar1 1],[\bar1 1 1]$ and $[1 \bar1 1]$. In the following, these four orientations will be denoted as NV$_{i}$, represented by the unit length vectors $\hat n_i$, $i \in \{1,2,3,4\}$ (see Fig.\:\ref{fig:fig1}(c)). 
In a sufficiently high external magnetic field, the magnetic field projection $B_\parallel$ along these four NV axes $B_i$ can be extracted from a total of eight resonances, in pairs of two due to transitions to the $\ket{m_S=-1}$ and $\ket{m_S=+1}$ spin states, in the ODMR signal as can be seen in Figure \ref{fig:fig1}(d).

However, the employment of this intrinsic tetrahedral coordinate system with the base vectors $\hat n_i$ for vector magnetometry is challenging due to the symmetries in $ \hat n_i$. To assign a pair of resonances to one of the NV axes, a known bias field is essential. 
By applying the bias field in an axis that is at a different angle to each of the four NV orientations, the NV axes NV$_{i}$ can be encoded with their angle to the bias field vector $\vartheta_i$. 
Previous implementations have used three-dimensional Helmholtz coils \cite{schloss2018simultaneous,zhao2019high,dmitriev2016concept,vershovskii2015micro}, permanent magnets \cite{davis2018mapping,wang2015high} or Halbach arrays \cite{clevenson2018robust,wickenbrock2021high} to apply the bias field, leading to a bulky and lab-bound setup and hindering the miniaturisation of the sensor device. 

In this paper, we present a novel approach where the bias field is generated by microscale wire coil wrapped around a diamond microcrystal that is placed on the tip of a multimode optical fiber (MMF) for optical access. This fiber-tip coil is simultaneously also used for spin manipulation at microwave frequencies. 
The sensor setup involves a two-step process: First, to employ the uniaxial bias field for vector magnetometry, the crystal orientation is defined in the bias field by preselection of diamond microcrystals via ODMR. However, with a single axis as reference, the exact crystal orientation remains unresolved in laboratory coordinates as there is still a rotational degree of freedom.
Hence, as a second step, the exact crystal orientation, namely the NV unit vectors $\hat{n}_i$, need to be determined in the laboratory coordinate system. 
As one of the fundamental challenges when using NV diamond particles such as diamond microcrystals or nanodiamonds for magnetic field sensing as opposed to bulk diamond, several methods have been proposed in the past to overcome this problem \cite{chen2019nitrogen, fukushige2020identification, li2023orientation, wang2023orientation}. 
Here, we use an approach in which a static, controlled magnetic field is scanned in three planes of rotation. Finally, we demonstrate fully fiber-integrated vector magnetometry in the full solid angle for static external magnetic fields with high dynamic range, as well as three-dimensional measurements of small and alternating changes in the external magnetic field. 
This development is a step towards robust and versatile application of three-dimensional, broadband, highly sensitive and spatially resolved NV magnetometry in confined spaces and extreme conditions.

\section{Sensor Setup}
\label{sec:methods}

As illustrated in Figure \ref{fig:fig1}(e), on the 50$\: \upmu\textrm{m}$ core of a cleaved multimode optical fiber with a numerical aperture $\textrm{NA} = 0.22$ (Thorlabs FG050UGA), a diamond microcrystal is positioned and fixed with optical adhesive (Norland NOA81). The diamond crystal has a size of $\sim15\: \upmu\textrm{m}$ and a NV$^-$ concentration of 3.5\:ppm and was supplied by Adamas Nanotechnologies (MDNV15umHi30mg). 
The fiber tip is positioned inside a copper wire coil that is used for simultaneously generation of MW and DC fields and is later described in detail. The two signals are generated in a signal generator (Rigol DSG836A) and a power supply in constant current mode (Rhode \& Schwarz NGE100), respectively, and combined in a bias tee (Minicircuits ZFBT-352-FT+). The signals are transmitted to the wire coil via coaxial cable. 

The diamond is optically addressed through the MMF by a $520\:\textrm{nm}$ diode laser that is reflected by a long pass dichroic mirror (Thorlabs DMLP550) and coupled to the fiber with a standard 10$\times$, $\textrm{NA} = 0.25$ microscope objective. 
The PL from the diamond passes through the dichroic mirror and a bandpass filterset with a cut-on wavelength of $550\:\textrm{nm}$ (Thorlabs FES0570) and a cut-off wavelength of $750\:\textrm{nm}$ (Thorlabs FES0750). It is focused onto an amplified PD (Thorlabs PDA26A2) that is set to a gain of $60\:\textrm{dB}$. 

The MW signal generated by the signal generator is modulated in either its amplitude (AM) or frequency (FM). The modulation frequency $f_{LF}$ is generated by a lock-in amplifier (LI) (Z\"urich Instruments MFLI), which also demodulates the PD signal at the same frequency. 
The LI output is then transmitted to a computer for data acquisition. This method effectively reduces noise, including 1/f noise, and is commonly employed in magnetometry with NV centers \cite{dix2024miniaturized,blakley2016fiber,fedotov2014electron,schloss2018simultaneous,clevenson2018robust,pogorzelski2024compact}. 
It is particularly advantageous when optical signals are transmitted through a fiber, as it mitigates artifacts such as fiber vibration and motion. In the following experiments, we primarily use AM modulation unless otherwise specified.

\subsection{Fiber Coupling of Pre-Selected Diamond Microcrystals}
\label{sec:fiber-coupling}

The goal of preselecting the diamond microcrystal for coupling to the fiber is to define the angles $\vartheta_i$ of the NV axes NV$_{i}$ in the bias field. Because the uniaxial bias field $\vec B_\textrm{bias}$ is fixed to the fiber axis is terms of orientation, we determine the diamond lattice orientation before fixating it to the fiber tip with optical adhesive. We see this technique of identifying single NV diamonds and picking them from a substrate with a fiber tip potentially being used for diamond of various sizes, e.g. nanodiamonds.

\begin{figure}[hb]
    \centering
    \includegraphics[width = 0.5\textwidth]{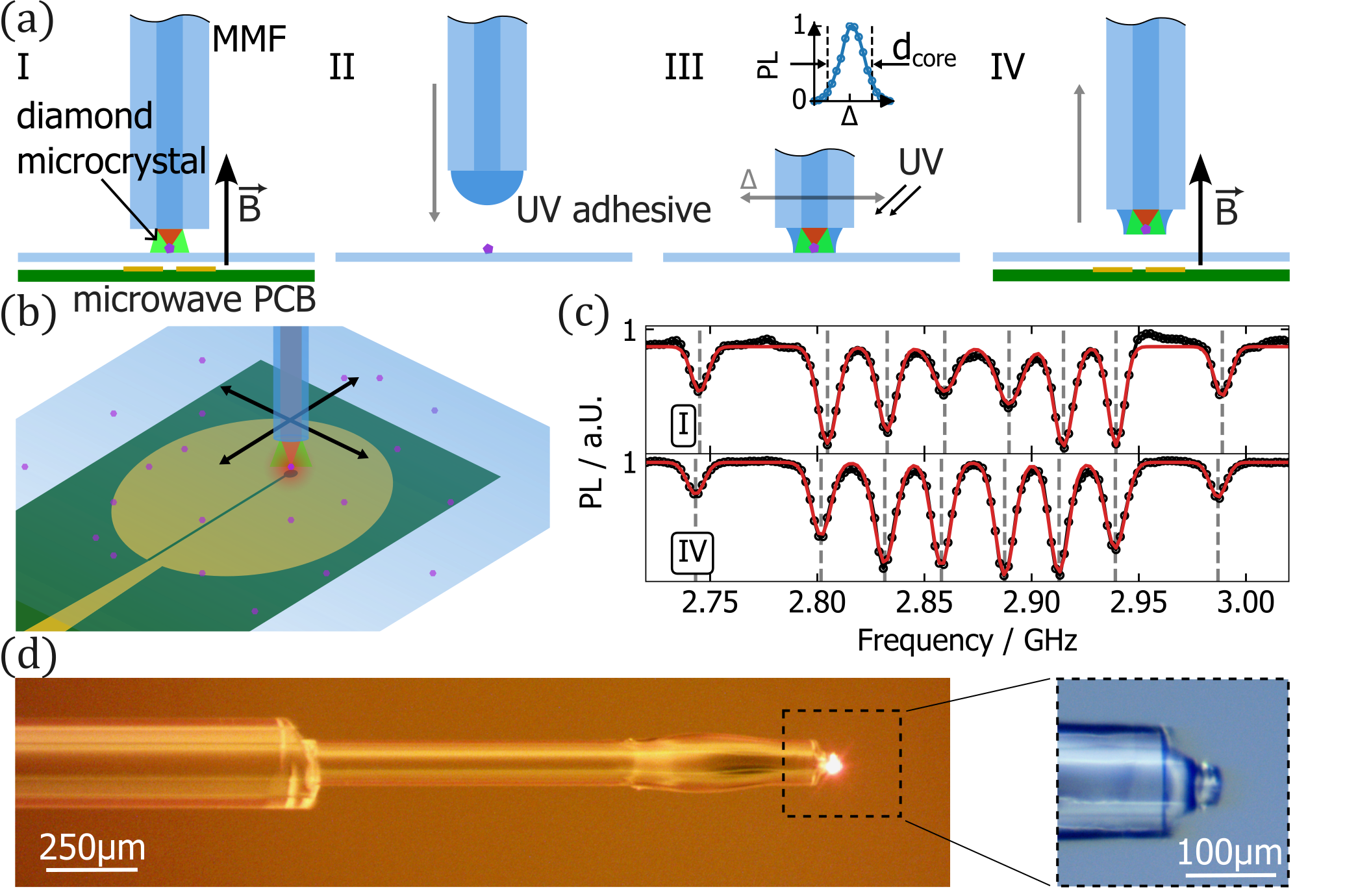}
    \caption{(a),(b) Individual diamond microcrystals containing NV centers are optically addressed through the MMF by scanning its tip laterally over the substrate. The crystal orientation is assessed by ODMR measurements in a magnetic field along the fiber axis with a PCB MW antenna \cite{Sasaki2016} (\texttt{I}). The diamond is fixed to the fiber tip with UV adhesive while the centering of the diamond microcrystal is monitored via the detected fluorescence (\texttt{II}, \texttt{III}, \texttt{IV}). (c) The separation of the resonances in the magnetic field along the fiber axis ensures that the NV centers can be labeled based on their respective angles in the bias field $\vartheta_i$ (top). After the fixation on the fiber tip (\texttt{IV}), the angles remain unchanged (bottom). (d) Diamond microcrystal with $15\:\upmu\textrm{m}$ diameter attached to the tip of a $d_\textrm{core} = 50\:\upmu\textrm{m}$ MMF with UV adhesive. The red NV fluorescence is clearly visible through a long pass filter ($550\:\textrm{nm}$ cut-off). }
    \label{fig:fig2}
\end{figure}

The diamond microcrystals are suspended in isopropyl alcohol and drop cast onto a glass substrate. After evaporation of the suspension liquid, individual diamond microcrystals can be identified by measuring the fluorescence through the MMF during lateral movement of the sample under the fiber tip (see Fig.\:\ref{fig:fig2}(b)). 
We acquire ODMR signals from the NV ensembles by manipulating the electron spin using a broadband PCB microwave resonator that has a sufficiently high field even at a few millimieters from the PCB \cite{Sasaki2016}. A custom-built Helmholtz coil is used to apply a magnetic field along the MMF, in the same axis as that generated by the fiber-tip wire coil. 
The main selection criterion is the measurement of eight frequency-separated resonances. This ensures that each of the four NV orientations can be distinguished by its angle to the bias field $\vec B_\mathrm{bias}$ generated by the fiber-tip coil. 

Once a diamond microcrystal with suitable orientation has been identified, a drop of optical UV adhesive is applied to the fiber tip, which is lowered until the liquid adhesive comes into contact with the particle. Lateral positioning can be fine-tuned by maximizing the fluorescence detected through the MMF (see inset of Fig.\:\ref{fig:fig2}(a)\texttt{III}). The adhesive is polymerized using a UV lamp to secure the position of the diamond microcrystals.

Figure \ref{fig:fig2}(c) shows ODMR measurements for the diamond microcrystal used in this work. From the measured ODMR signal during positioning (see Fig.\:\ref{fig:fig2}(a)\texttt{I}), we compute the angles $\vartheta_i$  of the NV axes NV$_{i}$ in the bias field $B_\mathrm{bias}$, as described in Appendix \ref{sec:appendix_evaluation}, and find $\vartheta_1 = 14.9\:^\circ,\:\vartheta_2 = 57.9\:^\circ,\:\vartheta_3 = 71.2\:^\circ$ and $\vartheta_4 = 83.6\:^\circ$. 
In the course of this, the ODMR signal is fitted with the sum of eight Gaussians in order to determine the resonance frequencies. Note that a traditional method is the fit with a Lorentzian profile \cite{graham2023fiber,chen2019nitrogen,chipaux2015magnetic}, however, in these experiments, fitting both profiles to the data showed that a Gaussian profile more closely resembled the measured data. 
We interpret this observation as inhomogeneous broadening of the lines which arises from different local environments of the individual NV spins \cite{barry2020sensitivity}.

In an ODMR measurement, after the diamond is fixed on the tip of the MMF, the angles are estimated as $\vartheta_1 = 15.5^\circ,\:\vartheta_2 = 57.3^\circ,\:\vartheta_3 = 71.5^\circ$ and $\vartheta_4 = 84.0^\circ$. This indicates a rotation of the diamond microcrystal of only $0.6\:^\circ$ during the positioning process. In our experiments, we did not observe any significant change of orientation of the diamond microcrystal unless the tip of the MMF comes into physical contact with the crystal. 

In the following, the four NV axes are denoted according to their angle in the bias field in ascending order, i.e. NV$_1$ represents the NV orientation with the smallest angle $\vartheta_1$ to the bias field $B_\mathrm{bias}$.

\subsection{Fiber-Integrated Magnetic Field Bias Coil}
\label{sec:coil}
\begin{figure}[ht]
\centering
  \includegraphics[width=0.5\textwidth]{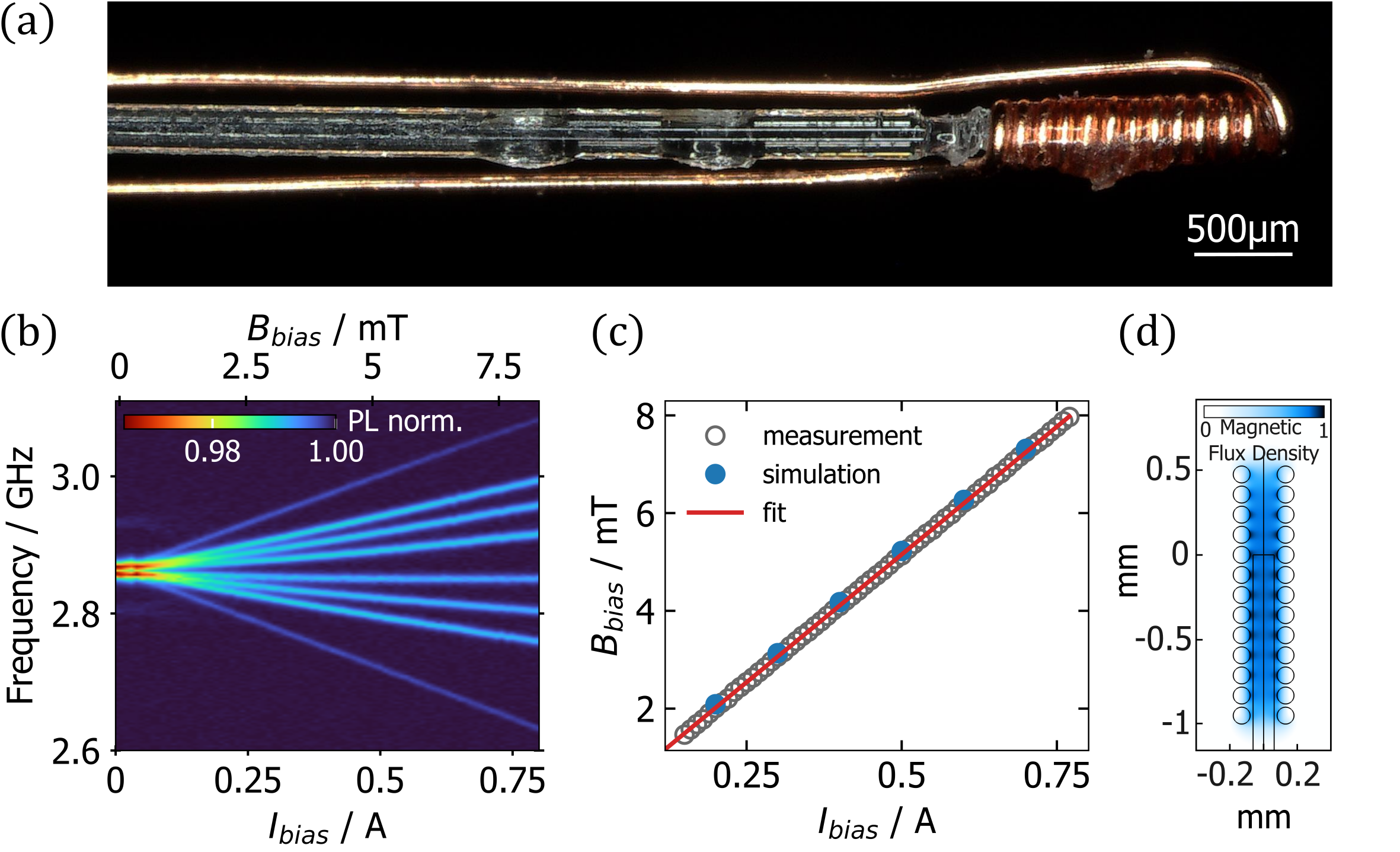}
  \caption{(a) Magnetic field bias coil on the fiber tip. The diamond microcrystal is positioned inside the coil. (b) ODMR measurements for different DC currents $I_\textrm{bias}$. In the low field regime $<1.5\:\textrm{mT}$, the degeneracy of the $\ket{m_S=\pm1}$ spin states is lifted due to internal crystal strain by $2E\approx11.9\:\textrm{MHz}$. (c) From these measurements, the magnetic field magnitude $B_\mathrm{bias}$ is deduced. Measurements in the low field regime are not taken into account due to the overlap of the resonances. The measured values are consistent with FEM simulations. (d) Simulated DC magnetic field distribution inside the fiber-tip coil.}
  \label{fig:fig3}
\end{figure}

For simultaneous generation of a microwave field at around $2.87\:\textrm{GHz}$ and a localized DC bias magnetic field in the diamond vicinity, a microscale wire coil is used (see Fig.\:\ref{fig:fig3}(a)). 
The single layer windings of this coil consist of a $100\:\upmu\textrm{m}$ diameter enamelled copper wire. It is manually wound around a $150\:\upmu\textrm{m}$ copper wire which is kept under tension to ensure straightness of the coil. The coil is then removed from the wire by cutting it and sliding the coil off with a pair of tweezers, using some isopropyl alcohol as a lubricant. 
The stripped MMF has a cladding diameter of $125\:\upmu\textrm{m}$ and can thus be inserted into the coil with a manual XYZ translation stage. The diamond is positioned inside the coil at the 9th of a total of 13 windings. The magnetic coil is finally secured with UV curing adhesive. 
The wire ends with a length of roughly $5\:\textrm{cm}$ are connected to the inner and outer conductor of a coaxial cable.

To calibrate the generated magnetic field of the wire coil, ODMR measurements are recorded while varying the current fed to the fiber-tip magnetic coil (see Fig.\:\ref{fig:fig3}(b)). The magnetic field magnitude is derived from nonlinear fitting as described in Appendix \ref{sec:appendix_evaluation}. In the low field regime, where $\gamma B_i \lesssim E$, the resonances overlap which leads to an inaccuracy of the fit. Therefore, we discard measurements for $I_\textrm{bias}<150\:\textrm{mA}$ and find a slope of $dB_\textrm{bias}/dI_\textrm{bias} = 10.46\:\textrm{mT/A}$, which is consistent with FEM simulations (see Fig.\:\ref{fig:fig3}(c)). Note that in our experiments we avoid $I_\textrm{bias}\geq 500\:\textrm{mA}$ in continuous operation to reduce current-induced heat generation at the fiber tip.

The FEM simulations were conducted with a wire diameter of $100\:\upmu\textrm{m}$ and an inner coil radius of $80\:\upmu\textrm{m}$ to account for the $5\:\upmu\textrm{m}$ film thickness of the isolation. 
The wire windings are spaced at a distance of $119\:\upmu\textrm{m}$ from center to center, considering imperfect spacing and film thickness of the isolation. The simulated field distribution is not entirely homogeneous along the fiber axis, as the coil diameter has similar dimensions to the wire diameter. 
However, considering the $15\:\upmu\textrm{m}$ size of the diamond crystal, the field distribution within the diamond volume can be assumed to be homogeneous. 
Inhomogeneous field distribution on the other hand would lead to an increase of linewidth with higher field magnitudes, since individual NV centers experience different local field strengths, which we did not observe in our measurements. 
Moreover, as evident from the simulation, the generated static field is well confined within the fiber-tip coil. This is very advantageous in situations where the measurement in an unconfined bias field would cause undesired interactions with the samples or devices under test.

\subsection{Determination of Crystal Orientation}
\label{sec:crystal_orientation}

The next step in the sensor setup is the determination of the NV axes in the laboratory coordinate system (herein lab frame). As stated in the introduction, the magnetic field components $B_i$ directed along the four NV axes $\hat{n}_i$, can be measured with ODMR. 
A crucial missing link is the $3\times4$ transformation matrix $\bm{K},$ that transforms the magnetic field vector $\vec B_{NV} = (B_1,B_2,B_3,B_4)$ in the non-orthogonal NV coordinate system with the unit vectors $\hat{n}_i$ (herein NV frame, denoted with $NV$ subscript) to the vector $\vec B$ in the lab frame,
\begin{equation}
    \vec B = \bm{K} \cdot \vec B_{NV}.
\label{eq:transformation}
\end{equation}
The columns of this matrix $\bm{K}$ simply contain the $x-,y-$ and $z-$ components of $\hat{n}_i$ in the lab frame. Note that once this matrix $\bm{K}$ is determined, the lab frame is fixed to the fiber tip, and it can be moved and rotated just like conventional 3D Hall sensors. To derive $\hat{n}_i$, we take ODMR measurements while scanning $\vec B$ in three planes of rotation.

\begin{figure*}[ht]
\centering
  \includegraphics[width=\textwidth]{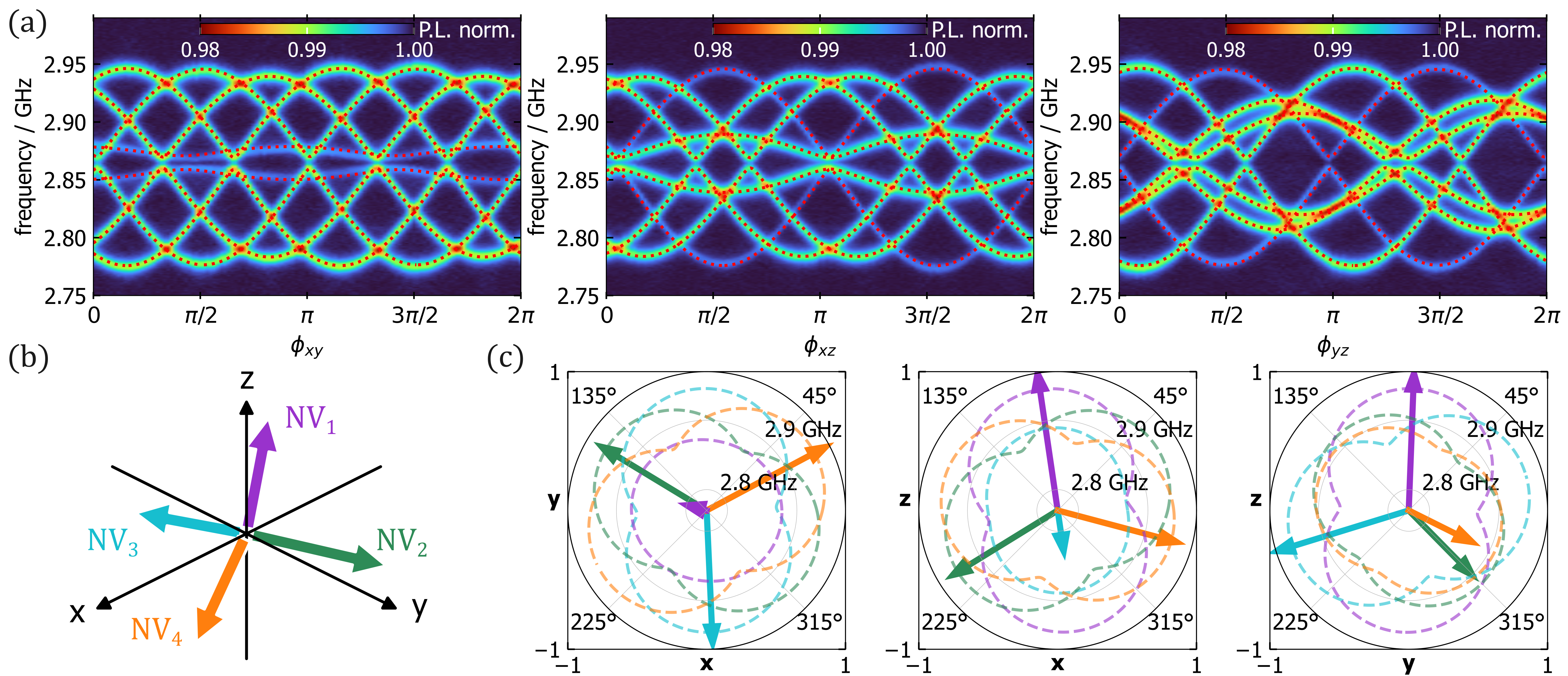}
  \caption{(a) ODMR measurements for a full rotation of the field vector with a magnitude of $B = 2.95\:\textrm{mT}$ in the $xy-$, $xz-$ and $yz-$plane (left to right). From the fit, depicted in dotted red lines, (b) the four NV axes NV$_{i}$ in unity vector representation are obtained. (c) The resulting NV axes, projected into the $xy-$, $xz-$ and $yz-$planes in cartesian coordinates. The transition frequencies to the $\ket{m_S = +1}$ spin states, as shown in (a), are depicted in polar coordinates for a full rotation of the field vector (dashed lines). The transition frequency corresponding to NV$_{i}$ in a given plane is maximized when the field vector is parallel or anti-parallel to the projection of $\hat n_i$ onto that plane (arrows).}
  \label{fig:fig4}
\end{figure*}

The fiber tip is positioned in the center of a custom-built 3D Helmholtz coil so that the fiber axis is aligned with $\vec e_{z}$ in the lab frame. ODMR measurements are taken from the fiber-coupled diamond microcrystal while the field vector is rotated about the azimuth angle $\phi$ so that 
\begin{equation}
  \begin{aligned}
    &\vec B_{xy} =B_0 \:(\cos \phi_{xy} ,\:\sin \phi_{xy} ,\: 0),\\
    &\vec B_{xz} =B_0\: (\cos \phi_{xz},\: 0 ,\:\sin \phi_{xz} )\text{ and}\\
    &\vec B_{yz} =B_0\: ( 0,\:\cos \phi_{yz} ,\:\sin \phi_{yz} ).
  \end{aligned}
  \label{eq:B_paramet}
\end{equation}

The results depicted in Figure \ref{fig:fig4}(a) illustrate the frequency-dependent normalized PL in pseudocolor across a full rotation of the magnetic field vector. Eight curves emerge in four pairs, corresponding to the spin transitions to the $\ket{m_S=\pm1}$ states for the four NV axes. 
The progression of resonances for a single NV axis for a complete rotation qualitatively resembles two half sine waves, with resonance frequency minima (maxima for $\ket{m_S =-1}$ transitions) occurring when $\vec B \perp \hat{n}_i$, and maxima (minima for $\ket{m_S = -1 }$ transitions) occurring when $\vec B$ is parallel or antiparallel to the projection of $\hat{n}_i$ onto the rotation plane (see Fig.\:\ref{fig:fig4}(c)). 
Consequently, the azimuth angle of each NV axis within the rotation plane can be inferred from the measurement. Furthermore, the elevation angle from the rotation plane determines the maximum resonance frequency (minimum for $\ket{m_S = -1}$ transitions) as it constrains the minimal angle between $\vec B$ and $\hat{n}_i$. 
However, since only the absolute value of $B_\parallel$ influences the transition frequency, the transition frequencies behave identically for both half rotations of the magnetic field vector. 
This introduces an ambiguity with respect to the sign of the elevation angle, specifically whether the NV axis is positioned above or below the reference plane, rendering it impossible to unambiguously determine the NV axes from a measurement in a single rotation plane.

As described in more detail in Appendix \ref{sec:appendix_rot_plane}, by parameterizing the resonance frequencies with the rotation angle $\phi$ derived from Eq.\:(\ref{eq:B_paramet}), we obtain a expression in which the resonance frequencies depend on $\phi$, the two in-plane components of the NV axes $\hat{n}_i$ (e.g. $n_x,n_y$ for rotation in the $xy-$plane), the external field magnitude $B_0$ and the constants of the NV ground state $\gamma, D$ and $E$. 
Thus, fitting this expression to the measured ODMR datasets lets us directly identify the in-plane components of $\hat{n}_i$ in each rotation plane. By defining the transformation matrix as 
\begin{equation}
    \bm{K} = \bm{R}\cdot \frac{1}{3}
    \begin{pmatrix}
        0 & 2\sqrt{2} & -\sqrt{2} & -\sqrt{2} \\
        0& 0& \sqrt{6}& -\sqrt{6} \\
        3 & -1 & -1 & -1
    \end{pmatrix} 
\end{equation}
where $\bm{R}=\bm{R}_x(\chi)\cdot\bm{R}_y(\psi)\cdot\bm{R}_z(\omega)$ are rotations around $\vec e_x,\vec e_y$ and $\vec e_z$ with the rotation angles $\chi,\psi$ and $\omega,$ we can optimize the rotation angles so that the root-mean-square error of the matrix elements of $\bm{K}$ and the vector components of $\hat{n}_i$ from the fit is minimized. 
We find $\chi=-2.19^\circ$, $\psi=-8.36^\circ$ and $\omega =149.67^\circ$. The vectors $\hat{n}_i$ depicted in Figure \ref{fig:fig4}(b) are represented as the columns of $\bm{K}$, where $\hat{n}_1$ is the first column and $\hat{n}_4$ is the last column of the matrix. 
Note that for Eq.\:(\ref{eq:transformation}) to be equal, a correction factor $c = \frac{3}{4}$ accounts for the extra element in $B_{NV}$ and the corrected matrix is $
    \bm{K_c} = \frac{3}{4} \bm{K}$.

\section{Measurements}

\subsection{Vectormagnetometry of Static Fields}

\begin{figure}[ht]
\centering
  \includegraphics[width=0.5\textwidth]{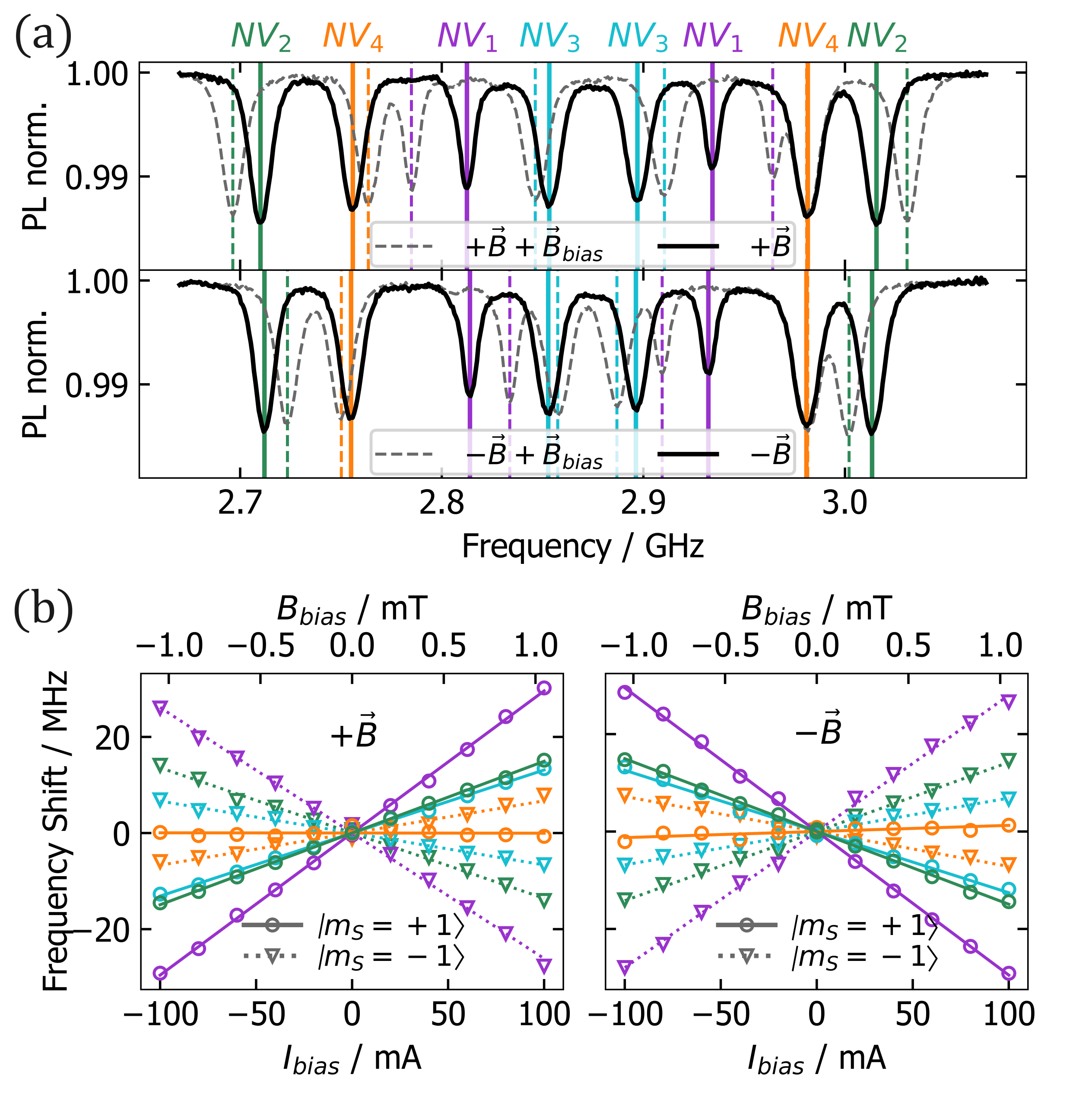}
  \caption{(a) Measured ODMR signal for a applied magnetic field $\vec B = (5.2,0,3)\:\textrm{mT}$ (upper) and $-\vec B = -(5.2,0,3)\:\textrm{mT}$ (lower). (b) The shift of the resonance frequencies is dependent on the orientation of the NV axes in the bias field $\vec B_\mathrm{bias}.$ Therefore, the frequency shift indicates the transition frequencies associated with each NV axis, as well as the field polarity.}
  \label{fig:fig5}
\end{figure}

To demonstrate NV vectormagnetometry, the fiberized NV-ensemble is placed in the center of a 3D Helmholtz coil for controlled field generation. A static magnetic field of $\vec B = (5.2,\:0,\:3)\:\textrm{mT}$ is applied. 
We measure a ODMR signal as displayed in Figure \ref{fig:fig5}. With no other information that the transition frequencies derived from the measurement, the assignment of the four NV axes to the transition frequencies is ambiguous. 
While this data is sufficient to determine the field magnitude, randomly assigning the NV axes to infer the azimuth angle $\varphi$ and polar angle $\vartheta$ would yield 24 possible permutations due to the inherent ambiguity. This count is doubled since both $+\vec B$ and $-\vec B$ have the same influence on the transition frequencies (see Fig.\:\ref{fig:fig5}(a)). 

To resolve these ambiguities, we introduce an additional field $\vec B_\textrm{bias}$ using the fiber-tip coil. Since the orientation of the NV ensemble in the bias field has been previously determined, the frequency shifts $\Delta f_i$ resulting from the bias field can be utilized to associate the resonances with the NV axes. 
Notably, the frequency shift $\Delta f_i$ depends on the angle $\vartheta_i$ between the NV axes $\hat{n}_i$ and the bias field $\vec B_\textrm{bias} = B_\textrm{bias} \vec e_z$. Specifically, according to the definition in Section \ref{sec:fiber-coupling}, the highest measured frequency shift corresponds to NV$_{1}$, followed by NV$_{2}$, NV$_{3}$, and NV$_{4}$ in descending order (see Fig.\:\ref{fig:fig5}).

From the measurements, the resonance frequencies are determined by a Gaussian fit. As shown in Appendix \ref{sec:appendix_evaluation}, from the resonance frequencies, we derive the magnetic field vector in the NV frame $\vec B_{NV} = (2.2,  -5.5,  -0.7, 4)\:\textrm{mT}$ and $\vec B_{NV} = (-2.1,  5.4,  -0.7, 4)\:\textrm{mT}$. 
The signs of the vector components are derived from the tetrahedral geometry of the NV ensemble and the direction of the frequency shift (see Fig.\:\ref{fig:fig5}(b)). The magnetic field in the lab coordinate system $\vec B$ is given by matrix multiplication with $\bm{K_c}$. The vectors $\vec B = (5.4,\:0.1,\:3.0)\:\textrm{mT}$ and $\vec B = (-5.4,\:-0.1,\:-2.9)\:\textrm{mT}$ can be determined with acceptable agreement with the applied magnetic field of $\vec B = (5.2\:,0\:,3)\:\textrm{mT}$ and $\vec B = (-5.2\:,0\:,-3)\:\textrm{mT}$. 
The respective angular discrepancies of $1.5^\circ$ and $2.0^\circ$ between the applied and the measured vectors can be attributed to a suboptimal centering of the NV diamond microcrystal in the Helmholtz coil. 

\subsection{Special Case of Overlapping Resonances}
In certain field configurations, components of $\vec B_{NV}$ may coincide, which has previously been considered as ``dead zones'' of the sensor, since the reconstruction of the applied field vector $\vec B$ is particularly challenging in these cases  \cite{vershovskii2015micro, dmitriev2016concept}. To demonstrate the absence of dead zones in the proposed setup, we apply a field such that the components $|B_1| = |B_2| = |B_3| = |B_4|$ of $\vec B_{NV}$ are equal, causing the resonances to overlap. Considering the tetrahedral geometry of the NV ensemble, the field vector can be oriented in six different directions to achieve equal absolute values for all components of $\vec B_{NV}$ as discussed in Appendix \ref{sec:appendix_deadzones}.

\begin{figure}[ht]
\centering
  \includegraphics[width=0.5\textwidth]{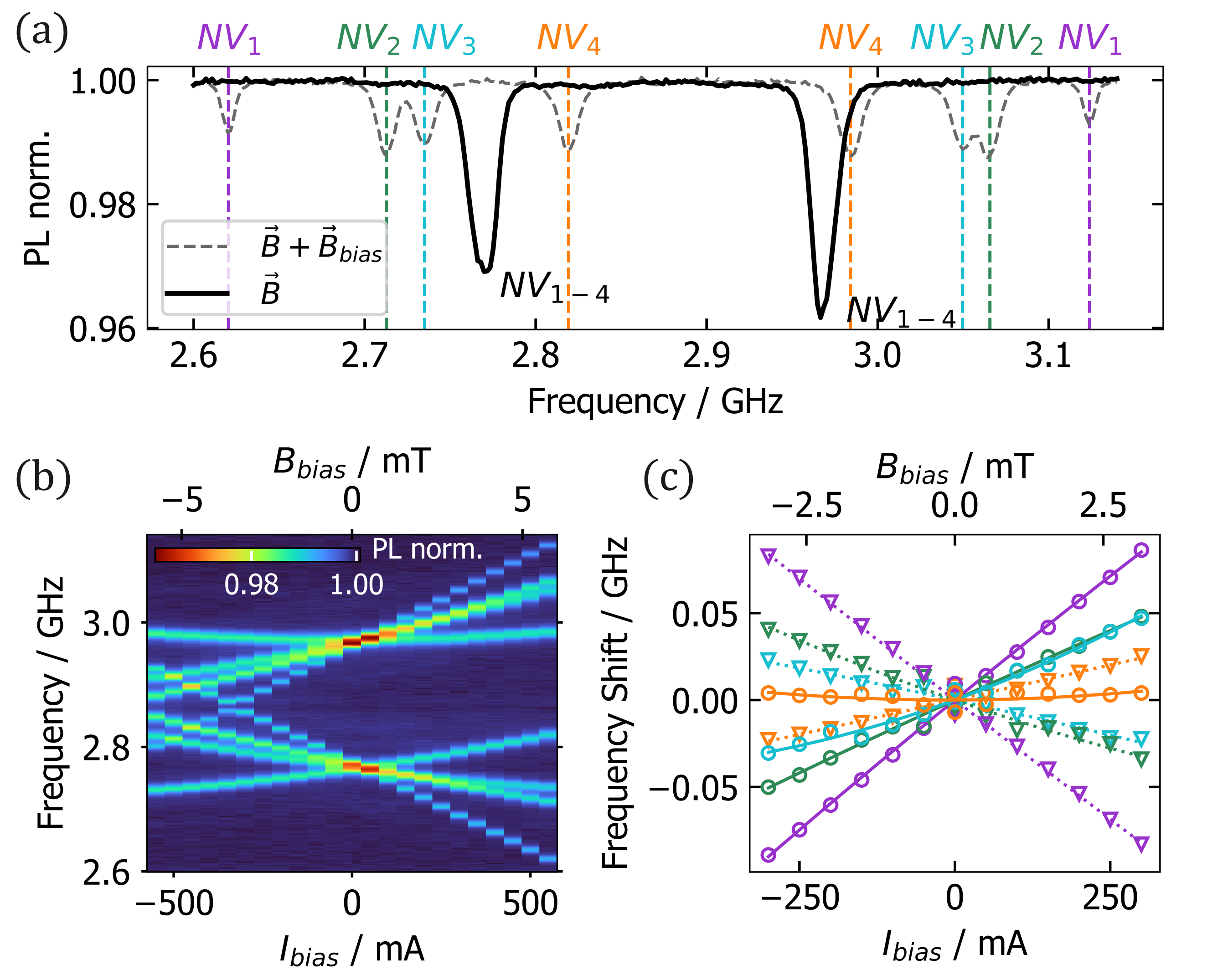}
  \caption{(a) ODMR measurement when $|B_1| = |B_2| = |B_3| = |B_4|.$ The magnetic field has a 6-fold ambiguity in this case in terms of its orientation. (b) The overlapping resonance lines can be seperated by applying a $B_\mathrm{bias}$ in the fiber-tip coil. (c) The direction of the frequency shifts $\Delta f_{i}$ corresponding to NV$_{i}$ has a signature behaviour for any of the six possible orientations of magnetic field.}
  \label{fig:fig6}
\end{figure}

When feeding a current to the fiber-tip coil, the resonances seperate as depicted in Figure \ref{fig:fig6}. The frequency shifts of the transitions to the $\ket{m_S=\pm 1}$ spin states exhibit distinct behavior. Notably, for at least one of the four NV orientations, the shift direction will invert for any of the six possible orientations of $\vec B$ as shown in Appendix \ref{sec:appendix_deadzones}. Thus, $\vec B$ can be unambiguously reconstructed from the measurement. We determine a vector $\vec B=(3.8,\: 2.6,\: 4.0)\:\textrm{mT}$, which deviates from the applied field $\vec B=(3.7,\: 2.6,\: 3.9)\:\textrm{mT}$ by a margin of $0.5^\circ$ in angular deviation. Note that while not being a dead zone, the sensor is less sensitive to angular changes in the external field for a case like this, as the resonances will overlap in a frequency range that corresponds to the linewidth of the transitions. 

\subsection{Vectormagnetometry of Dynamic Fields by Frequency Modulated ODMR}
\label{sec:fm_odmr}
\begin{figure}[ht]
\centering
  \includegraphics[width=0.5\textwidth]{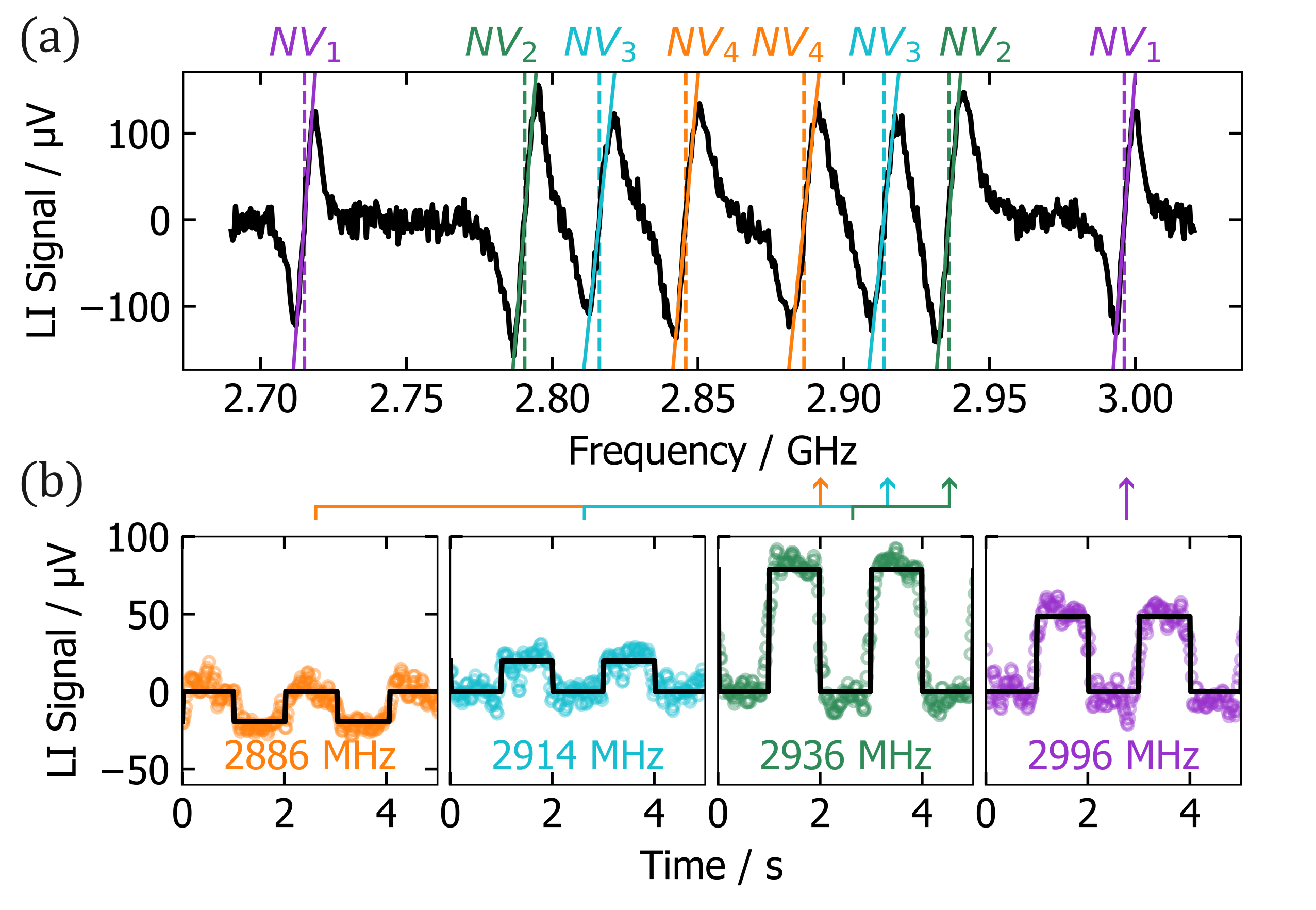}
  \caption{(a) FM-ODMR measurement in a bias field with the fiber-tip coil of $B_\textrm{bias}=5.2\:\textrm{mT}$. At the frequencies  the zero crossings of the derivatives of the resonances, the LI signal is magnetically sensitive and small changes in $B_{i}$ along NV$_{i}$ can be directly read out. (b) 4$\times$ averaged time traces of the demodulated signal in each frequency band when applying a square wave external magnetic field with an amplitude of $B=70.7\:\upmu\textrm{T}$ and a frequency of $f=0.5\:\textrm{Hz}.$}
  \label{fig:fig7}
\end{figure}

While the measurement of the magnetic field via a sweep of MW frequency has a high dynamic range up to several tens of milliteslas \cite{steinert2010high,homrighausen2023edge,nowodzinski2015nitrogen}, it relies on the external field being fully static during the acquistion time. 
In order to retrieve real time measurements, methods have been presented in the past that utilize frequency modulation (FM) of the microwave frequency in order to individually interrogate ODMR features in a bias field of several mT \cite{graham2023fiber,schloss2018simultaneous,clevenson2018robust,pogorzelski2024compact,sturner2021integrated}. 
When sweeping a frequency modulated microwave frequency, the demodulated signal is a derivative of the lineshapes that would be acquired from CW-ODMR acquisition techniques as can be seen in Figure \ref{fig:fig7}(a). At the frequency of the zero crossing, the LI signal is a directly related to changes in the magnetic field, as it scales with the gyromagnetic ratio $\gamma$, the component of the external field $B_{i}$ along the respective NV axis NV$_{i}$ and the slope of the zero crossing in units of $\mathrm{V/Hz}$. 
Thus, a small change in the external field $\vec B$ will detune the resonance frequency which is detected in a increase or a decrease of the LI signal. This method enables real-time measurements of the magnetic field vector $\vec B$ when performed in parallel \cite{schloss2018simultaneous} or scalar fields when only one transition frequency is interrogated \cite{pogorzelski2024compact,sturner2021integrated}, which is limited in its bandwidth only by the modulation frequency of the FM \cite{clevenson2018robust}.

In order to demonstrate the three-dimensional measurement of dynamic changes in the magnetic fields, a square wave signal with a period of $T=2\:\mathrm{s}$ is applied to the $x-$ and $z-$ of the 3D Helmholtz coils in series, which generates an external magnetic field with an amplitude of $\vec B = (49.9,0,50.4) \:\mathrm{\upmu T}$. 
The MW frequency is modulated with a modulation depth of $f_\Delta = 500 \:\mathrm{kHz}$ and a modulation frequency of $f_{LF} = 3 \:\mathrm{kHz}$. For each of the four transitions to the $\ket{m_S=+1}$ spin states, the time trace of the LI signal is acquired sequentially over 18 seconds. Note that sequential readout introduces a dead time in each frequency band and therefore relies on a periodic signal, whereas simultaneous readout, as demonstrated in \cite{schloss2018simultaneous}, allows the measurement of non-periodic signals.
Figure \ref{fig:fig7}(b) shows the offset corrected four times average of the LI signal with a filter bandwidth of $5\:\textrm{Hz}$. By fitting a 11th-order fourier series of a square wave signal to the data, the amplitudes of the LI signal at each frequency is obtained. 
With the gyromagnetic ratio $\gamma$ and the slope at the zero crossing, derived from Figure \ref{fig:fig7}(a), the resulting vector in the NV frame is $\vec B_{NV} = (38.1,-67.5,-20.9,20.8)\:\upmu\textrm{T}$. 
Here, the sign of each component $B_{i}$ is derived from the direction of the frequency shift and the known orientations of NV$_{i}$ and $\vec B_\textrm{bias}$. Matrix multiplication with $\bm{K}$ yields $\vec B = (46.9,0,52.9)\: \upmu\textrm{T}$ which is in good agreement with the applied external field vector. 
The deviation from the expected values may arise from a noisy signal and a suboptimal centering of the fiber tip in the 3D Helmholtz coils.

\subsection{Fiber Tip Sensor Characteristics}
\label{sec:sensor_characteristics}

A main figure of merits for a magnetic sensing device is the magnetic sensitivity. Here, we assess the shot-noise limited sensitivity which is a theoretical lower limit when neglecting all technical noise sources. It is calculated for a single resonance with 
\begin{equation}
    \eta_B = P_G \frac{1}{\gamma} \frac{\Delta \nu}{C \sqrt{R}},
\end{equation}
where $P_G = \sqrt{e/8\ln 2}$ is a factor for a gaussian lineshape, $\Delta \nu$ is the full width half maximum (FWHM) linewidth, $C$ is the ODMR contrast of the resonance and $R$ is the detected count rate \cite{dreau2011avoiding}. 
It has been shown that for NV centers, both linewidth and contrast depend on the MW field strength, with both quantities generally increasing with rising field strength \cite{dreau2011avoiding}. 
However, the broadening of the resonance lines becomes more pronounced with high MW powers, outweighing the increase in ODMR contrast in terms of magnetic sensitivity. This counteracting behaviour leads to a region of MW field strength where the shot-noise limited sensitivity is minimized. 
There are also reports of decreasing linewidth with very efficient optical pumping \cite{jensen2013light}, which we did not observe in our experiments. 

To find the optimum in sensitivity, the MW power $P_{MW}$ of the signal generator is varied while maintaining a constant laser power $P_{L}=100\:\textrm{mW}.$ From the ODMR measurements in a bias field, we fit eight Gaussians to the data, as described in Appendix \ref{sec:appendix_evaluation}, and consequently obtain the contrast and linewidth of all eight resonance lines. In other implementations of NV magnetometers where only single orientations of NV centers are interrogated, it may be purposeful to optimize single resonances and only estimate the sensitivity of that resonance, i.e. via the orientation of the respective NV axis in the MW field. Here on the other hand, since the magnetic fields along all four NV axes $B_i$ are observed, we average the ODMR contrast and the linewidth and find a optimum of shot-noise limited sensitivity of $\eta_B = 19.4\:\textrm{nT/Hz}^{1/2}$ for $P_{MW} = 12.5\:\textrm{dBm}$ with a count rate $R=5.9\times10^{12}$, a contrast of $C=1.2\%$ and a FWHM linewidth of $\Delta \nu = 11.9\:\textrm{MHz}.$

\begin{figure}[ht]
\centering
  \includegraphics[width=0.45\textwidth]{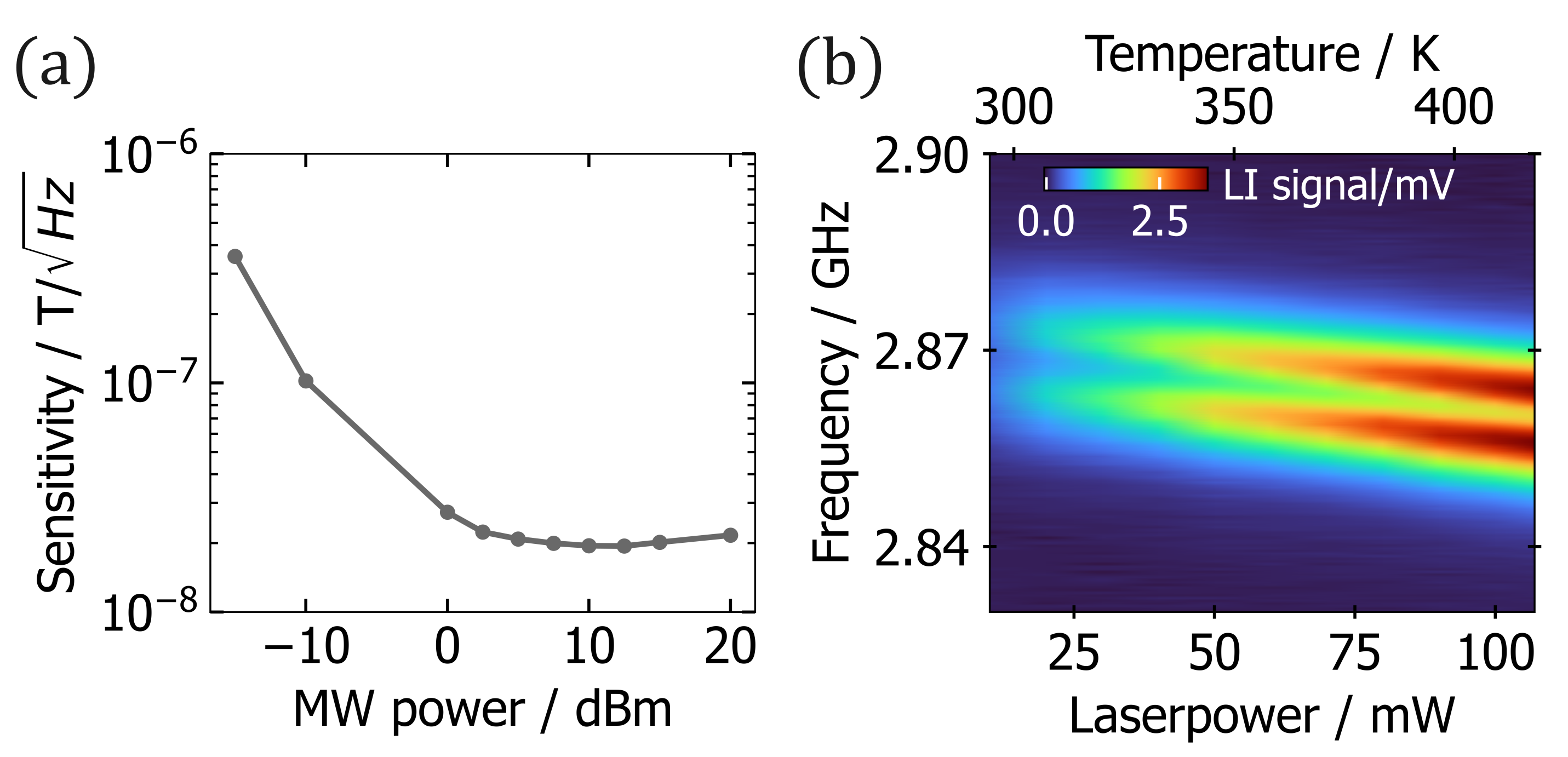}
  \caption{(a) Estimated shot-noise-limited sensitivity of the NV magnetometer with a laser power of $P_{L}=100\:\textrm{mW}$. With an optimum of $\eta_B = 19.4\:\textrm{nT/Hz}^{1/2}$ at $P_{MW}=12.5\:\textrm{dBm},$ the sensitivity degrades for higher values of $P_{MW}$ due to power broadening of the resonances. (b) Zero field ODMR measurements for different values of $P_L$. The degeneracy of the $\ket{m_S=\pm1}$ is lifted due to internal crystal strain by $2E=11.9\:\textrm{MHz}.$ From the shift of the zero field splitting parameter $D=2.87\:\textrm{GHz},$ the diamond temperature for $P_{L}=100\:\textrm{mW}$ is estimated to be $109.7\:\textrm{K}$ above room temperature.}
  \label{fig:fig8}
\end{figure}

Furthermore, we evaluate the temperature at the fiber tip, which can be observed by a shift of the zero field splitting parameter $D=2.87\:\textrm{GHz}$ towards lower frequencies with a coefficent of $dD/dT=-74.2\:\textrm{kHz/K}$ \cite{acosta2010temperature}. 
This is done by varying the laser power during zero-field ODMR measurements, as shown in Figure \ref{fig:fig8}(b). At a laser power of $P_L=100\:\textrm{mW},$ a shift of $\Delta D=8.14\:\textrm{MHz}$ is measured which corresponds to a temperature increase of $\Delta T = 109.7\:\textrm{K}.$ 
We find a linear dependence of the temperature on the laser power, and did not observe any additional heating effects from the MW power or the current flowing through the fiber-tip coil. Even at higher currents up to $I_c=500\:\textrm{mA},$ the diamond temperature remains stable over time. 
In addition, within the investigated range of laser power, the fluorescence scales linearly with the laser power which leads to the conclusion that the intensity at the fiber tip is still far below the saturation intensity of the NV diamond microcrystal. 

The laser heating of the diamond can be partially attributed to the mismatch between the diamond microcrystal size of $15\:\upmu\textrm{m}$ to the core diameter of the multimode fiber $50\:\upmu\textrm{m}$ and a resulting suboptimal optical pumping efficiency of the NV centers.
Although heating effects are known to counteract a straightforward increase of pump light power to achieve higher count rates and consequently better sensitivities in integrated diamond sensors \cite{pogorzelski2024compact,dix2024miniaturized}, we expect that a better matching of the diamond size to the fiber core diameter should reduce the requirement for high laser powers and thus improve the usability of the fiber sensor in temperature critical applications e.g. life science. 
Furthermore, the fluorescence collection efficiency is currently limited by the numerical aperture $\textrm{NA}=0.22$ of the MMF and may be improved by using a MMF with a higher numerical aperture and implementing additional optical elements on the fiber facet. With a reliable control over the diamond temperature, the presented sensor design could also be used for temperature sensing.

\section{Conclusion and Outlook}

In conclusion, we have presented a fully fiber-integrated vector magnetic field sensor based on NV ensembles by the generation of a localized bias field on the fiber tip with a fiber-tip coil. 
A major advancement is the usage of this fiber-tip coil is used simultaneously for the bias field generation and the spin control by combining DC and MW signals with microwave frequency in a bias tee. 

A novel preselection method is implemented to utilize the one-dimensional bias field for vectormagnetometry, in the course of which the orientation of the diamond microcrystal is interrogated. 
Subsequently, the NV axes of the ensemble are accurately determined by data sets of ODMR measurements while scanning a controlled static field in three planes of rotation, and a matrix is obtained for the transformation from the four-element vector in the NV frame to the three-dimensional vector in laboratory coordinates. 

Active field control using coils, as opposed to a field bias with permanent magnets, allows versatile vectormagnetometry in the full solid angle. We demonstrate the measurement of static fields with a high dynamic range by sweeping the microwave frequency and then use a small bias field to unambiguously assign the resonances to the NV axes of the ensemble. 
By the vector addition of the bias field and the external field, certain dead zones that arise from the overlap of resonances due to the geometry of the diamond crystal lattice can be resolved. Moreover, using a high bias field in the range of several milliteslas enables measurements of AC signals with a limited dynamic range by detecting small changes in the magnetic field along the four NV axes. 
As an outlook, separating the detection channels in the frequency domain for parallel readout would enhance the acquisition time and enable three-dimensional measurements of arbitrarily shaped signals. Potential applications of the presented fiber sensor could include power monitoring and efficiency enhancement in electric motors, wind turbines, transformers or photovoltaic inverters by providing precise measurements of vector magnetic fields within narrow gaps, thus contributing to advancements in the energy transition.

For the presented sensor, we estimate a shot-noise limited magnetic sensitivity of $19.4\:\textrm{nT/Hz}^{1/2}$ with a sensor cross section below $1\:\textrm{mm}^2$ and a spatial resolution of $15\:\upmu\textrm{m}$. 
A limiting factor is the detected count rate and laser heating of the diamond which can be improved by matching the diamond size to the core diameter of the optical fiber, increasing the fiber NA or by using additional optical elements to increase the pumping efficiency as well as the fluorescence collection efficiency. 
Further improvements in terms of sensitivity include the optimization of the diamond samples purity, which would lead to a reduced linewidth at comparable microwave powers.

\begin{acknowledgments}
This research was funded by the German Federal Ministry of Education and Research under the project OCQNV (grant number 13N15971) and by the Ministry of Culture and Science of the State of North-Rhine Westfalia under the project E2QNV (grant number 005-2302-0021). We thank Ludwig Horsthemke and the other members of the HLB laboratory of the FH M\"unster for fruitful discussions and support with the driver circuits for the 3D Helmholtz coil.

The authors declare no conflict of interest.

Data underlying the results presented in this paper are not publicly available at this time but may be obtained from the authors upon reasonable request.

This version of the manuscript was recently accepted as a research article in \href{https://journals.aps.org/prapplied/}{APS Physical Review Applied}.

\end{acknowledgments}

\appendix

\section{Zeeman Shift of the Electron Spin Levels}
\label{sec:appendix_transf}
In this section, we describe the behaviour of the resonance frequencies $f_\pm$ in order to derive an expression used for the evaluation of the measurements. The contents of this section are closely adapted from reference \cite{doherty2012theory}. The approximate eigenstates of the ground state spin-Hamiltionian, expressed as frequencies, are 
\begin{equation}
    f_\pm = D + \frac{(\gamma B_\perp)^2}{D} \pm \xi \sqrt{ (\gamma B_\parallel)^2 + E^2 } 
    \label{eq:trans_f}
\end{equation}
with the dimensionless correction factor 
\begin{equation*}
    \xi = \sqrt{ 1 +  \frac{\gamma^4 B_\perp^4}{ 4D^2( \gamma^2 B_\parallel^2 + E^2 )} }
\end{equation*}
where $D$ is the zero field splitting parameter, $\gamma$ is the gyromagnetic ratio, $B_\parallel$ and $B_\perp$ are the axial and non-axial magnetic fields and $E$ is the non-axial electric-strain field parameter. 
The axial electric-strain field parameter $E_\parallel$ is included in $D$ as in $D = D'+E_\parallel$, since in our experiments, $D$ is temperature shifted and therefore must be calibrated as the effective zero field splitting parameter $D$. 
Furthermore, a term in $\xi$ is omitted that includes the expression $\cos (2\phi_B+\phi_E)$ where, in a local coordinate system in which $\vec e_z$ is directed along the NV axes, $ \phi_B =\arctan (B_x/B_y)$ and $\phi_E = \arctan (E_x/E_y)$ are the azimuth angles of the non-axial magnetic and electric-strain field components. 
Since the strain in diamond is believed to arise from local impurities and dislocations \cite{barry2020sensitivity}, we assume $\phi_E$ to be random for every NV center in the ensemble which results in $ \frac{1}{n} \sum_i^n \cos (2\phi_{B,i}+\phi_{E,i}) \approx 0$ when averaged over $n$ NV centers.

\section{Evaluation of Magnetic Field Magnitude and Field Angles}
\label{sec:appendix_evaluation}
To find field information from ODMR measurements such as the field magnitude, the field angles in the NV frame $\vartheta_i$ and the field vector in the NV frame $\vec B$ first, the resonance frequencies of NV$_{i}$ must be determined. For this, the sum of eight Gaussian profiles is fitted to the measured normalized ODMR data in the form of 
\begin{equation}
    G(f) = 1 - \sum_i^8  C_i e^{\left({-4 \ln{2} \left(\frac{f-f_{\pm,i}}{\Delta \nu_i}\right)^2} \right)},
\end{equation}
where $C$ is the ODMR contrast, $\Delta \nu$ is the FWHM line width and $f_\pm$ are the resonance frequencies.

We assume a locally defined coordinate system $x'  y' z'$ in the crystal lattice, in which the $z$-Axis is oriented along $\hat n_1$ and the $y$-component of $\hat n_2$ is zero. In this coordinate system, the matrix $\bm{K'}$ that transform $\vec B_{NV}$ according to 
\begin{equation}
    \vec B'= \bm{K'} \cdot \vec B_{NV}
\end{equation}
is 
\begin{equation}
    \bm{K'} = \frac{1}{3}
    \begin{pmatrix}
        0 & 2\sqrt{2} & -\sqrt{2} & -\sqrt{2} \\
        0& 0& \sqrt{6}& -\sqrt{6} \\
        3 & -1 & -1 & -1
    \end{pmatrix}.
    \label{eq:kmatrix}
\end{equation}
The columns of the matrix $\bm{K'}$ correspond to the unit length NV axes in vector representation $\hat n_i$.
This local coordinate system $ x'  y' z'$ is related to the laboratory coordinate system $ x  y  z$ via the rotation matrix $\bm{R}$:
\begin{equation}
     \bm{K} = \bm{R} \cdot \bm{K'}.
     \label{eq:rotmatrix}
\end{equation}

For a given magnetic field vector in spherical coordinates $\vec B=(B_0,\phi,\theta)$, all eight resonance frequencies $f_{\pm,i}$ are calculated from Eq.\:(\ref{eq:trans_f}). 
Here, the axial field component $B_{\parallel,i}$ and the non-axial field component $B_{\perp_,i}$ for each NV axis NV$_{i}$ are calculated via $B_{\parallel,i}=B_0 \cos{\vartheta_i}$ and $B_{\perp,i}=B_0 \sin{\vartheta_i},$ where $\cos \vartheta_i = (\hat n_i \cdot \vec B)/B_0$. 
With the constants of the NV ground state $D,E,$ and $\gamma$, we can optimize the parameters $B_0$, $\phi$ and $\theta$ of the resulting expression $f_{\pm,i}(B_0,\phi,\theta)$ to the resonance frequencies derived from the fit of the Gaussians to the measured data to determine $\vec B$. 
Note that $D$ has to be calibrated for a given laser power as it shifts with temperature \cite{acosta2010temperature} and $E$ is calibrated once for the diamond microcrystal, as it depends on the local crystal strain \cite{dolde2011electric}. 
For optimization, either nonlinear least-squares fitting or minimization of the root mean square error with dual annealing \cite{xiang1997generalized} is used, where the latter seemed to perform better with discontinuous functions.

However, if the transformation matrix $\bm{R}$ is unknown, the resulting vector $\vec B$ is defined in $\hat x' \hat y' \hat z'$ and only the magnetic magnitude $B_0$ as well as the angles $\vartheta_i$ relative to $\hat n_i$ can be derived from this vector, where the angles $\vartheta_i$ are shuffled randomly when the resonance frequencies are not assigned to the NV axes in the right order. Also, the polarity of $\vec B'$ remains unresolved.

\section{Finding the Rotation Matrix from ODMR in Planes of Rotation}
\label{sec:appendix_rot_plane}
To describe the behaviour of the resonance frequencies while rotating the magnetic field vector $\vec B$ in a plane of rotation, we need to parameterise Eq.\:(\ref{eq:trans_f}) with the rotation angle $\phi.$ 

Exemplary for a rotation of $\vec B$ in the $xy-$plane, the magnetic field vector is 
\begin{equation}
    \vec B = \begin{pmatrix}
        B_0 \cos{\phi} \\ 
        B_0 \sin{\phi} \\
        0
    \end{pmatrix}.
\end{equation}
For a given NV axes $\hat n$, the axial component $B_\parallel$ and the non-axial component $B_\perp$ are 
\begin{equation}
    B_\parallel(\phi) = \hat n \cdot \vec{B} = B_0( n_x \cos{\phi} + n_y \sin{\phi} ) 
\end{equation} and 
\begin{equation}
    B_\perp(\phi) = \sqrt{B_0^2 - B_\parallel^2} = B_0 \sqrt{1 - (n_x \cos{\phi} + n_y \sin{\phi})^2} 
\end{equation}
where $n_x$ and $n_y$ are the $x-$ and $y-$component of $\hat n$. The term $(n_x \cos{\phi} + n_y \sin{\phi})$ will be referred to as $n_{ij}(\phi)$, where $i$ and $j$ are the in-plane components of $\hat n$.

With these expressions, according to (\ref{eq:trans_f}) the resonance frequencies for a rotation in the $ij-$plane are given by 

\begin{equation}
    f_\pm(\phi)= D + \frac{(\gamma B_0)^2}{D}  (1-n_{ij}^2(\phi)) \pm \xi \sqrt{ \left( \gamma B_0 n_{ij}(\phi) \right)^2 + E^2 }
    \label{eq:trans_f2}
\end{equation}
where 
\begin{equation*}
     \xi = \sqrt{ 1 + \frac{\gamma^4 B_0^4 (1-n_{ij}^2(\phi))^2 }{4 D^2  \left[\left( \gamma B_0 n_{ij}(\phi) \right)^2 + E^2\right] } } \:\:.
\end{equation*}
In this representation, all parameters are known except $n_{ij}$ which are the in-plane components of $\hat n$. Thus, doing the measurements in different rotation planes, namely the $xy-$, $xz-$ and $yz-$planes, the complete vector $\hat n$ can be determined.

To fit Eq.\:(\ref{eq:trans_f2}) to the measured data, a sum of eight Gaussians is fitted to the data via nonlinear optimization as described in Appendix \ref{sec:appendix_evaluation} to extract the resonance frequencies. 
The result is a $N\times8$ dimensional array of resonance frequencies where $N$ is the number of rotation angles in the dataset, making it non-trivial to fit a model function in the form of $f_\pm=f(\phi)$ to the data. 

Hence, we employ an algorithm that is very similar to Dijkstra's path finding \cite{dijkstra1959note}. First, all resonance frequencies are reshaped to a point set with the point coordinates $p=(x,y) = (\phi,f_\pm).$ 
Secondly, from a starting point $p_0=(x_0,y_0)$, weights are assigned to all points with $y>y_0$ based on their Euclidean distance $d_1=\sqrt{(x-x_0)^2 + (y-y_0)^2}$ to $p_0$ and their distance $d_2= \sqrt{(x-x_0)^2 + (f(x)-y_0)^2}$ to a model function $f(x)$. The weight of each point is $w_p = \sqrt{\left(c_1\frac{\min(d_1)}{d_{1,p}}\right)^2 + \left(c_2\frac{\min(d_2)}{d_{2,p}}\right) ^2}$, where $\min(d_1)$ and $\min(d_2)$ is the minimum of the resulting values for all points. 
The highest weighted point is appended to the path and will be treated as $p_0$ in the next iteration. The constants $c_1$ and $c_2$ are used to fine tune the performance of the algorithm. In our experiments, for units of $\textrm{radiants}$ for the rotation angle and units of megahertz for the frequency, $c_1 = c_2$ showed good results. 
Because of computational cost, we use a model function $f(x) = a + b |\sin(x-c)|$ for the algorithm, which closely resembles Eq.\:(\ref{eq:trans_f2}) but is less complex. The parameters of this model function are fitted to the existing path after each iteration of this algorithm. 
By setting reasonable bounds to the parameters $a,b$ and $c$ for the fit and repeating the algorithms for different starting points, we are able to find eight distinct paths through the point set corresponding to the $\ket{m_S=\pm1}$ spin transitions for each of the four NV symmetry axes. 

Thirdly, Eq.\:(\ref{eq:trans_f2}) is fitted to each path, and we directly obtain the in-plane components of $\hat n_i$ from the fit, e.g. $n_x$ and $n_y$ for rotation in the $xy-$plane. The results od this fit are depicted in the main text in Figure \ref{fig:fig4}(a) in red dotted lines. 
The out-of-plane component, e.g. $n_z$ for rotation in the $xy-$plane, can be retrieved from the in-plane components and the unit-length nature of the vector, however, in our experiments, the out-of-plane component has a notably higher standard deviation compared to the in-plane components, which is why it is set to zero. 
Consequently, from rotation of $\vec B$ in the three planes, we obtain a total of 24 vectors, two for the $\ket{m_S=\pm1}$ spin transitions for each of the four NV symmetry axes.

Lastly, we define a rotation matrix $\bm{R} = \bm{R}_x(\chi)\bm{R}_y(\psi)\bm{R}_z(\omega)$ where $\bm{R}_x(A)$ ($\bm{R}_y(B)$, $\bm{R}_z(\Gamma)$) is a rotation around $\vec e_x$ ($\vec e_y$, $\vec e_z$) with an angle of $\chi$ ($\psi$, $\omega$). 
This matrix $\bm{R}$ transforms the local coordinate system $ x'  y'  z'$ relative to the crystal lattice to the laboratory coordinate system $ x  y  z$, as defined in Eq.\:(\ref{eq:kmatrix}). To find the angles $\chi$,$\psi$ and $\omega$, we multiply the matrices according to (\ref{eq:rotmatrix}) and treat these vectors as predictions, while treating the vectors given by the fit as observables. 
By minimizing the quadratic mean of the residuals of each vector component (root-mean-square error) with dual annealing \cite{xiang1997generalized}, the angles $\chi$, $\psi$ and $\omega$ of the rotation matrix $\bm{R}$ are derived.

\section{\label{sec:appendix_deadzones}Overlap of the Resonances}

The overlap of the resonances originate from equal field projections along the four NV axes $|B_1| = |B_2| = |B_3| = |B_4|$. This condition is met only when $\vec B$ lies in the plane spanned by two NV axes, while simultaneously lying in the perpendicular plane spanned by the other two NV axes. 
In this configuration, two components of $\vec B_{NV}$, corresponding to those spanning one of the planes, must be negative. This problem is a permutation with repetition and we can find the amount of possible vectors $\vec B$ by calculating $4! / (2!\cdot 2!) = 6.$ 
\begin{figure}[ht]
    \centering
    \includegraphics[width = 0.5\textwidth]{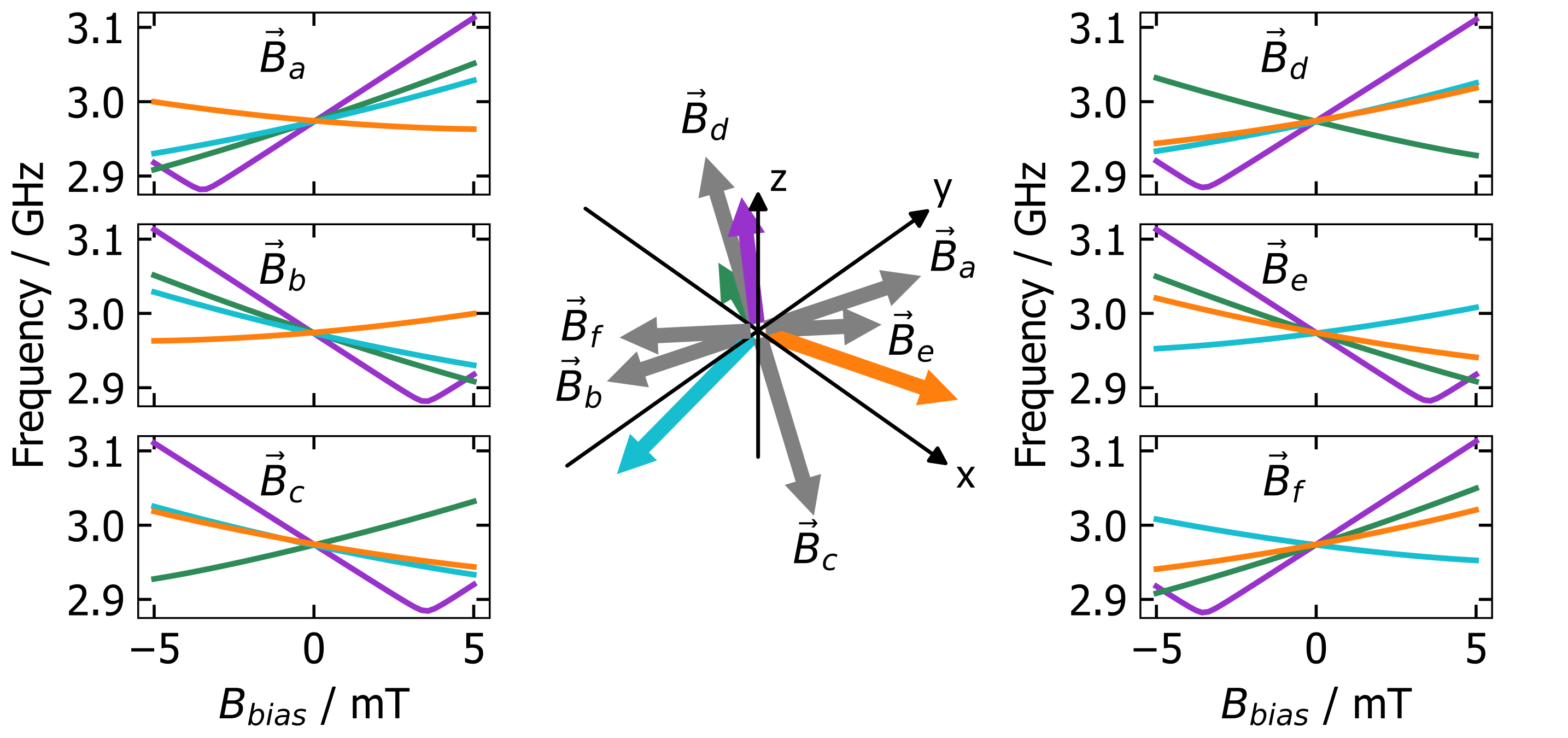}
    \caption{All cases of $\vec B$ (gray) where the projections of $\vec B$ onto the NV axes (colored) are equal. The resonance frequencies of the $\ket{m_S=+1}$ spin states, associated with NV$_{i}$, are depicted on the left and right with varying values the magnitude of $\vec B_\textrm{bias} = -B_\textrm{bias} \vec e_z$ and $B=6\:\textrm{mT}$. The resonance frequencies show distinct behaviour for every of the six cases of $\vec B$.}
    \label{fig:appendix2}
\end{figure}
The vectors are depicted in Figure \ref{fig:appendix2}, along with the expected frequencies of the $\ket{m_S=0}\rightarrow \ket{m_S=+1}$ spin transitions when applying a bias field anti-parallel to the $z-$ axis. These exhibit a distinct behavior, which is due to the vector addition of $\vec B + \vec B_\textrm{bias}$ giving different results for different $\vec B$. 
When comparing Figure \ref{fig:appendix2} to the measurement in the main text (Fig.\:\ref{fig:fig6}), it is evident that the vector $\vec B_a $ was applied in the experiment.

\bibliography{Bibliothek}

\begin{thebibliography}{48}%
\makeatletter
\providecommand \@ifxundefined [1]{%
 \@ifx{#1\undefined}
}%
\providecommand \@ifnum [1]{%
 \ifnum #1\expandafter \@firstoftwo
 \else \expandafter \@secondoftwo
 \fi
}%
\providecommand \@ifx [1]{%
 \ifx #1\expandafter \@firstoftwo
 \else \expandafter \@secondoftwo
 \fi
}%
\providecommand \natexlab [1]{#1}%
\providecommand \enquote  [1]{``#1''}%
\providecommand \bibnamefont  [1]{#1}%
\providecommand \bibfnamefont [1]{#1}%
\providecommand \citenamefont [1]{#1}%
\providecommand \href@noop [0]{\@secondoftwo}%
\providecommand \href [0]{\begingroup \@sanitize@url \@href}%
\providecommand \@href[1]{\@@startlink{#1}\@@href}%
\providecommand \@@href[1]{\endgroup#1\@@endlink}%
\providecommand \@sanitize@url [0]{\catcode `\\12\catcode `\$12\catcode
  `\&12\catcode `\#12\catcode `\^12\catcode `\_12\catcode `\%12\relax}%
\providecommand \@@startlink[1]{}%
\providecommand \@@endlink[0]{}%
\providecommand \url  [0]{\begingroup\@sanitize@url \@url }%
\providecommand \@url [1]{\endgroup\@href {#1}{\urlprefix }}%
\providecommand \urlprefix  [0]{URL }%
\providecommand \Eprint [0]{\href }%
\providecommand \doibase [0]{http://dx.doi.org/}%
\providecommand \selectlanguage [0]{\@gobble}%
\providecommand \bibinfo  [0]{\@secondoftwo}%
\providecommand \bibfield  [0]{\@secondoftwo}%
\providecommand \translation [1]{[#1]}%
\providecommand \BibitemOpen [0]{}%
\providecommand \bibitemStop [0]{}%
\providecommand \bibitemNoStop [0]{.\EOS\space}%
\providecommand \EOS [0]{\spacefactor3000\relax}%
\providecommand \BibitemShut  [1]{\csname bibitem#1\endcsname}%
\let\auto@bib@innerbib\@empty
\bibitem [{\citenamefont {Tetienne}\ \emph {et~al.}(2014)\citenamefont
  {Tetienne}, \citenamefont {Hingant}, \citenamefont {Kim}, \citenamefont
  {Diez}, \citenamefont {Adam}, \citenamefont {Garcia}, \citenamefont {Roch},
  \citenamefont {Rohart}, \citenamefont {Thiaville}, \citenamefont
  {Ravelosona},\ and\ \citenamefont {Jacques}}]{tetienne2014nanoscale}%
  \BibitemOpen
  \bibfield  {author} {\bibinfo {author} {\bibfnamefont {J.-P.}\ \bibnamefont
  {Tetienne}}, \bibinfo {author} {\bibfnamefont {T.}~\bibnamefont {Hingant}},
  \bibinfo {author} {\bibfnamefont {J.-V.}\ \bibnamefont {Kim}}, \bibinfo
  {author} {\bibfnamefont {L.~Herrera}\ \bibnamefont {Diez}}, \bibinfo {author}
  {\bibfnamefont {J.-P.}\ \bibnamefont {Adam}}, \bibinfo {author}
  {\bibfnamefont {K.}~\bibnamefont {Garcia}}, \bibinfo {author} {\bibfnamefont
  {J.-F.}\ \bibnamefont {Roch}}, \bibinfo {author} {\bibfnamefont
  {S.}~\bibnamefont {Rohart}}, \bibinfo {author} {\bibfnamefont
  {A.}~\bibnamefont {Thiaville}}, \bibinfo {author} {\bibfnamefont
  {D.}~\bibnamefont {Ravelosona}}, \ and\ \bibinfo {author} {\bibfnamefont
  {V.}~\bibnamefont {Jacques}},\ }\bibfield  {title} {\enquote {\bibinfo
  {title} {Nanoscale imaging and control of domain-wall hopping with a
  nitrogen-vacancy center microscope},}\ }\href {\doibase DOI:
  10.1126/science.1250113} {\bibfield  {journal} {\bibinfo  {journal}
  {Science}\ }\textbf {\bibinfo {volume} {344}},\ \bibinfo {pages} {1366--1369}
  (\bibinfo {year} {2014})}\BibitemShut {NoStop}%
\bibitem [{\citenamefont {Kucsko}\ \emph {et~al.}(2013)\citenamefont {Kucsko},
  \citenamefont {Maurer}, \citenamefont {Yao}, \citenamefont {Kubo},
  \citenamefont {Noh}, \citenamefont {Lo}, \citenamefont {Park},\ and\
  \citenamefont {Lukin}}]{kucsko2013nanometre}%
  \BibitemOpen
  \bibfield  {author} {\bibinfo {author} {\bibfnamefont {G.}~\bibnamefont
  {Kucsko}}, \bibinfo {author} {\bibfnamefont {P.~C.}\ \bibnamefont {Maurer}},
  \bibinfo {author} {\bibfnamefont {N.~Y.}\ \bibnamefont {Yao}}, \bibinfo
  {author} {\bibfnamefont {M.}~\bibnamefont {Kubo}}, \bibinfo {author}
  {\bibfnamefont {H.~J.}\ \bibnamefont {Noh}}, \bibinfo {author} {\bibfnamefont
  {P.~K.}\ \bibnamefont {Lo}}, \bibinfo {author} {\bibfnamefont
  {H.}~\bibnamefont {Park}}, \ and\ \bibinfo {author} {\bibfnamefont {M.~D.}\
  \bibnamefont {Lukin}},\ }\bibfield  {title} {\enquote {\bibinfo {title}
  {Nanometre-scale thermometry in a living cell},}\ }\href {\doibase
  https://doi.org/10.1038/nature12373} {\bibfield  {journal} {\bibinfo
  {journal} {Nature}\ }\textbf {\bibinfo {volume} {500}},\ \bibinfo {pages}
  {54--58} (\bibinfo {year} {2013})}\BibitemShut {NoStop}%
\bibitem [{\citenamefont {Wang}\ \emph {et~al.}(2021)\citenamefont {Wang},
  \citenamefont {Liu}, \citenamefont {Zhu},\ and\ \citenamefont
  {Cappellaro}}]{wang2021nanoscale}%
  \BibitemOpen
  \bibfield  {author} {\bibinfo {author} {\bibfnamefont {G.}~\bibnamefont
  {Wang}}, \bibinfo {author} {\bibfnamefont {Y.-X.}\ \bibnamefont {Liu}},
  \bibinfo {author} {\bibfnamefont {Y.}~\bibnamefont {Zhu}}, \ and\ \bibinfo
  {author} {\bibfnamefont {P.}~\bibnamefont {Cappellaro}},\ }\bibfield  {title}
  {\enquote {\bibinfo {title} {Nanoscale vector ac magnetometry with a single
  nitrogen-vacancy center in diamond},}\ }\href {\doibase
  https://doi.org/10.1021/acs.nanolett.1c01165} {\bibfield  {journal} {\bibinfo
   {journal} {Nano Letters}\ }\textbf {\bibinfo {volume} {21}},\ \bibinfo
  {pages} {5143--5150} (\bibinfo {year} {2021})}\BibitemShut {NoStop}%
\bibitem [{\citenamefont {Schloss}\ \emph {et~al.}(2018)\citenamefont
  {Schloss}, \citenamefont {Barry}, \citenamefont {Turner},\ and\ \citenamefont
  {Walsworth}}]{schloss2018simultaneous}%
  \BibitemOpen
  \bibfield  {author} {\bibinfo {author} {\bibfnamefont {J.~M.}\ \bibnamefont
  {Schloss}}, \bibinfo {author} {\bibfnamefont {J.~F.}\ \bibnamefont {Barry}},
  \bibinfo {author} {\bibfnamefont {M.~J.}\ \bibnamefont {Turner}}, \ and\
  \bibinfo {author} {\bibfnamefont {R.~L.}\ \bibnamefont {Walsworth}},\
  }\bibfield  {title} {\enquote {\bibinfo {title} {Simultaneous broadband
  vector magnetometry using solid-state spins},}\ }\href {\doibase
  https://doi.org/10.1103/PhysRevApplied.10.034044} {\bibfield  {journal}
  {\bibinfo  {journal} {Physical Review Applied}\ }\textbf {\bibinfo {volume}
  {10}},\ \bibinfo {pages} {034044} (\bibinfo {year} {2018})}\BibitemShut
  {NoStop}%
\bibitem [{\citenamefont {Shao}\ \emph {et~al.}(2016)\citenamefont {Shao},
  \citenamefont {Liu}, \citenamefont {Zhang}, \citenamefont {Shneidman},
  \citenamefont {Audier}, \citenamefont {Markham}, \citenamefont {Dhillon},
  \citenamefont {Twitchen}, \citenamefont {Xiao},\ and\ \citenamefont
  {Lon{\v{c}}ar}}]{shao2016wide}%
  \BibitemOpen
  \bibfield  {author} {\bibinfo {author} {\bibfnamefont {L.}~\bibnamefont
  {Shao}}, \bibinfo {author} {\bibfnamefont {R.}~\bibnamefont {Liu}}, \bibinfo
  {author} {\bibfnamefont {M.}~\bibnamefont {Zhang}}, \bibinfo {author}
  {\bibfnamefont {A.~V.}\ \bibnamefont {Shneidman}}, \bibinfo {author}
  {\bibfnamefont {X.}~\bibnamefont {Audier}}, \bibinfo {author} {\bibfnamefont
  {M.}~\bibnamefont {Markham}}, \bibinfo {author} {\bibfnamefont
  {H.}~\bibnamefont {Dhillon}}, \bibinfo {author} {\bibfnamefont {D.~J.}\
  \bibnamefont {Twitchen}}, \bibinfo {author} {\bibfnamefont {Y.-F.}\
  \bibnamefont {Xiao}}, \ and\ \bibinfo {author} {\bibfnamefont
  {M.}~\bibnamefont {Lon{\v{c}}ar}},\ }\bibfield  {title} {\enquote {\bibinfo
  {title} {Wide-field optical microscopy of microwave fields using
  nitrogen-vacancy centers in diamonds},}\ }\href {\doibase
  https://doi.org/10.1002/adom.201600039} {\bibfield  {journal} {\bibinfo
  {journal} {Advanced optical materials}\ }\textbf {\bibinfo {volume} {4}},\
  \bibinfo {pages} {1075--1080} (\bibinfo {year} {2016})}\BibitemShut {NoStop}%
\bibitem [{\citenamefont {Graham}\ \emph {et~al.}(2023)\citenamefont {Graham},
  \citenamefont {Rahman}, \citenamefont {Munn}, \citenamefont {Patel},
  \citenamefont {Newman}, \citenamefont {Stephen}, \citenamefont {Colston},
  \citenamefont {Nikitin}, \citenamefont {Edmonds}, \citenamefont {Twitchen},
  \citenamefont {Markham},\ and\ \citenamefont {Morley}}]{graham2023fiber}%
  \BibitemOpen
  \bibfield  {author} {\bibinfo {author} {\bibfnamefont {S.~M.}\ \bibnamefont
  {Graham}}, \bibinfo {author} {\bibfnamefont {A.~T. M.~A.}\ \bibnamefont
  {Rahman}}, \bibinfo {author} {\bibfnamefont {L.}~\bibnamefont {Munn}},
  \bibinfo {author} {\bibfnamefont {R.~L.}\ \bibnamefont {Patel}}, \bibinfo
  {author} {\bibfnamefont {A.~J.}\ \bibnamefont {Newman}}, \bibinfo {author}
  {\bibfnamefont {C.~J.}\ \bibnamefont {Stephen}}, \bibinfo {author}
  {\bibfnamefont {G.}~\bibnamefont {Colston}}, \bibinfo {author} {\bibfnamefont
  {A.}~\bibnamefont {Nikitin}}, \bibinfo {author} {\bibfnamefont {A.~M.}\
  \bibnamefont {Edmonds}}, \bibinfo {author} {\bibfnamefont {D.~J.}\
  \bibnamefont {Twitchen}}, \bibinfo {author} {\bibfnamefont {M.~L.}\
  \bibnamefont {Markham}}, \ and\ \bibinfo {author} {\bibfnamefont {G.~W.}\
  \bibnamefont {Morley}},\ }\bibfield  {title} {\enquote {\bibinfo {title}
  {Fiber-coupled diamond magnetometry with an unshielded sensitivity of 30
  pt/hz},}\ }\href {\doibase https://doi.org/10.1103/PhysRevApplied.19.044042}
  {\bibfield  {journal} {\bibinfo  {journal} {Physical Review Applied}\
  }\textbf {\bibinfo {volume} {19}},\ \bibinfo {pages} {044042} (\bibinfo
  {year} {2023})}\BibitemShut {NoStop}%
\bibitem [{\citenamefont {Wolf}\ \emph {et~al.}(2015)\citenamefont {Wolf},
  \citenamefont {Neumann}, \citenamefont {Nakamura}, \citenamefont {Sumiya},
  \citenamefont {Ohshima}, \citenamefont {Isoya},\ and\ \citenamefont
  {Wrachtrup}}]{wolf2015subpicotesla}%
  \BibitemOpen
  \bibfield  {author} {\bibinfo {author} {\bibfnamefont {T.}~\bibnamefont
  {Wolf}}, \bibinfo {author} {\bibfnamefont {P.}~\bibnamefont {Neumann}},
  \bibinfo {author} {\bibfnamefont {K.}~\bibnamefont {Nakamura}}, \bibinfo
  {author} {\bibfnamefont {H.}~\bibnamefont {Sumiya}}, \bibinfo {author}
  {\bibfnamefont {T.}~\bibnamefont {Ohshima}}, \bibinfo {author} {\bibfnamefont
  {J.}~\bibnamefont {Isoya}}, \ and\ \bibinfo {author} {\bibfnamefont
  {J.}~\bibnamefont {Wrachtrup}},\ }\bibfield  {title} {\enquote {\bibinfo
  {title} {Subpicotesla diamond magnetometry},}\ }\href {\doibase
  https://doi.org/10.1103/PhysRevX.5.041001} {\bibfield  {journal} {\bibinfo
  {journal} {Physical Review X}\ }\textbf {\bibinfo {volume} {5}},\ \bibinfo
  {pages} {041001} (\bibinfo {year} {2015})}\BibitemShut {NoStop}%
\bibitem [{\citenamefont {Barry}\ \emph {et~al.}(2016)\citenamefont {Barry},
  \citenamefont {Turner}, \citenamefont {Schloss}, \citenamefont {Glenn},
  \citenamefont {Song}, \citenamefont {Lukin}, \citenamefont {Park},\ and\
  \citenamefont {Walsworth}}]{barry2016optical}%
  \BibitemOpen
  \bibfield  {author} {\bibinfo {author} {\bibfnamefont {J.~F.}\ \bibnamefont
  {Barry}}, \bibinfo {author} {\bibfnamefont {M.~J.}\ \bibnamefont {Turner}},
  \bibinfo {author} {\bibfnamefont {J.~M.}\ \bibnamefont {Schloss}}, \bibinfo
  {author} {\bibfnamefont {D.~R.}\ \bibnamefont {Glenn}}, \bibinfo {author}
  {\bibfnamefont {Y.}~\bibnamefont {Song}}, \bibinfo {author} {\bibfnamefont
  {M.~D.}\ \bibnamefont {Lukin}}, \bibinfo {author} {\bibfnamefont
  {H.}~\bibnamefont {Park}}, \ and\ \bibinfo {author} {\bibfnamefont {R.~L.}\
  \bibnamefont {Walsworth}},\ }\bibfield  {title} {\enquote {\bibinfo {title}
  {Optical magnetic detection of single-neuron action potentials using quantum
  defects in diamond},}\ }\href {\doibase
  https://doi.org/10.1073/pnas.1601513113} {\bibfield  {journal} {\bibinfo
  {journal} {Proceedings of the National Academy of Sciences}\ }\textbf
  {\bibinfo {volume} {113}},\ \bibinfo {pages} {14133--14138} (\bibinfo {year}
  {2016})}\BibitemShut {NoStop}%
\bibitem [{\citenamefont {Xie}\ \emph {et~al.}(2021)\citenamefont {Xie},
  \citenamefont {Yu}, \citenamefont {Zhu}, \citenamefont {Qin}, \citenamefont
  {Rong}, \citenamefont {Duan},\ and\ \citenamefont {Du}}]{xie2021hybrid}%
  \BibitemOpen
  \bibfield  {author} {\bibinfo {author} {\bibfnamefont {Y.}~\bibnamefont
  {Xie}}, \bibinfo {author} {\bibfnamefont {H.}~\bibnamefont {Yu}}, \bibinfo
  {author} {\bibfnamefont {Y.}~\bibnamefont {Zhu}}, \bibinfo {author}
  {\bibfnamefont {X.}~\bibnamefont {Qin}}, \bibinfo {author} {\bibfnamefont
  {X.}~\bibnamefont {Rong}}, \bibinfo {author} {\bibfnamefont {C.-K.}\
  \bibnamefont {Duan}}, \ and\ \bibinfo {author} {\bibfnamefont
  {J.}~\bibnamefont {Du}},\ }\bibfield  {title} {\enquote {\bibinfo {title} {A
  hybrid magnetometer towards femtotesla sensitivity under ambient
  conditions},}\ }\href {\doibase https://doi.org/10.1016/j.scib.2020.08.001}
  {\bibfield  {journal} {\bibinfo  {journal} {Science Bulletin}\ }\textbf
  {\bibinfo {volume} {66}},\ \bibinfo {pages} {127--132} (\bibinfo {year}
  {2021})}\BibitemShut {NoStop}%
\bibitem [{\citenamefont {Fescenko}\ \emph {et~al.}(2020)\citenamefont
  {Fescenko}, \citenamefont {Jarmola}, \citenamefont {Savukov}, \citenamefont
  {Kehayias}, \citenamefont {Smits}, \citenamefont {Damron}, \citenamefont
  {Ristoff}, \citenamefont {Mosavian},\ and\ \citenamefont
  {Acosta}}]{fescenko2020diamond}%
  \BibitemOpen
  \bibfield  {author} {\bibinfo {author} {\bibfnamefont {I.}~\bibnamefont
  {Fescenko}}, \bibinfo {author} {\bibfnamefont {A.}~\bibnamefont {Jarmola}},
  \bibinfo {author} {\bibfnamefont {I.}~\bibnamefont {Savukov}}, \bibinfo
  {author} {\bibfnamefont {P.}~\bibnamefont {Kehayias}}, \bibinfo {author}
  {\bibfnamefont {J.}~\bibnamefont {Smits}}, \bibinfo {author} {\bibfnamefont
  {J.}~\bibnamefont {Damron}}, \bibinfo {author} {\bibfnamefont
  {N.}~\bibnamefont {Ristoff}}, \bibinfo {author} {\bibfnamefont
  {N.}~\bibnamefont {Mosavian}}, \ and\ \bibinfo {author} {\bibfnamefont
  {V.~M.}\ \bibnamefont {Acosta}},\ }\bibfield  {title} {\enquote {\bibinfo
  {title} {Diamond magnetometer enhanced by ferrite flux concentrators},}\
  }\href {\doibase https://doi.org/10.1103/PhysRevResearch.2.023394} {\bibfield
   {journal} {\bibinfo  {journal} {Physical Review Research}\ }\textbf
  {\bibinfo {volume} {2}},\ \bibinfo {pages} {023394} (\bibinfo {year}
  {2020})}\BibitemShut {NoStop}%
\bibitem [{\citenamefont {Kim}\ \emph {et~al.}(2019)\citenamefont {Kim},
  \citenamefont {Ibrahim}, \citenamefont {Foy}, \citenamefont {Trusheim},
  \citenamefont {Han},\ and\ \citenamefont {Englund}}]{kim2019cmos}%
  \BibitemOpen
  \bibfield  {author} {\bibinfo {author} {\bibfnamefont {D.}~\bibnamefont
  {Kim}}, \bibinfo {author} {\bibfnamefont {M.~I.}\ \bibnamefont {Ibrahim}},
  \bibinfo {author} {\bibfnamefont {C.}~\bibnamefont {Foy}}, \bibinfo {author}
  {\bibfnamefont {M.~E.}\ \bibnamefont {Trusheim}}, \bibinfo {author}
  {\bibfnamefont {R.}~\bibnamefont {Han}}, \ and\ \bibinfo {author}
  {\bibfnamefont {D.~R.}\ \bibnamefont {Englund}},\ }\bibfield  {title}
  {\enquote {\bibinfo {title} {A cmos-integrated quantum sensor based on
  nitrogen--vacancy centres},}\ }\href {\doibase
  https://doi.org/10.1038/s41928-019-0275-5} {\bibfield  {journal} {\bibinfo
  {journal} {Nature Electronics}\ }\textbf {\bibinfo {volume} {2}},\ \bibinfo
  {pages} {284--289} (\bibinfo {year} {2019})}\BibitemShut {NoStop}%
\bibitem [{\citenamefont {Pogorzelski}\ \emph {et~al.}(2024)\citenamefont
  {Pogorzelski}, \citenamefont {Horsthemke}, \citenamefont {Homrighausen},
  \citenamefont {Stiegek{\"o}tter}, \citenamefont {Gregor},\ and\ \citenamefont
  {Gl{\"o}sek{\"o}tter}}]{pogorzelski2024compact}%
  \BibitemOpen
  \bibfield  {author} {\bibinfo {author} {\bibfnamefont {J.}~\bibnamefont
  {Pogorzelski}}, \bibinfo {author} {\bibfnamefont {L.}~\bibnamefont
  {Horsthemke}}, \bibinfo {author} {\bibfnamefont {J.}~\bibnamefont
  {Homrighausen}}, \bibinfo {author} {\bibfnamefont {D.}~\bibnamefont
  {Stiegek{\"o}tter}}, \bibinfo {author} {\bibfnamefont {M.}~\bibnamefont
  {Gregor}}, \ and\ \bibinfo {author} {\bibfnamefont {P.}~\bibnamefont
  {Gl{\"o}sek{\"o}tter}},\ }\bibfield  {title} {\enquote {\bibinfo {title}
  {Compact and fully integrated led quantum sensor based on nv centers in
  diamond},}\ }\href {\doibase https://doi.org/10.3390/s24030743} {\bibfield
  {journal} {\bibinfo  {journal} {Sensors}\ }\textbf {\bibinfo {volume} {24}},\
  \bibinfo {pages} {743} (\bibinfo {year} {2024})}\BibitemShut {NoStop}%
\bibitem [{\citenamefont {Toyli}\ \emph {et~al.}(2012)\citenamefont {Toyli},
  \citenamefont {Christle}, \citenamefont {Alkauskas}, \citenamefont {Buckley},
  \citenamefont {Van~de Walle},\ and\ \citenamefont
  {Awschalom}}]{toyli2012measurement}%
  \BibitemOpen
  \bibfield  {author} {\bibinfo {author} {\bibfnamefont {D.~M.}\ \bibnamefont
  {Toyli}}, \bibinfo {author} {\bibfnamefont {D.~J.}\ \bibnamefont {Christle}},
  \bibinfo {author} {\bibfnamefont {A.}~\bibnamefont {Alkauskas}}, \bibinfo
  {author} {\bibfnamefont {B.~B.}\ \bibnamefont {Buckley}}, \bibinfo {author}
  {\bibfnamefont {C.~G.}\ \bibnamefont {Van~de Walle}}, \ and\ \bibinfo
  {author} {\bibfnamefont {D.~D.}\ \bibnamefont {Awschalom}},\ }\bibfield
  {title} {\enquote {\bibinfo {title} {Measurement and control of single
  nitrogen-vacancy center spins above 600 k},}\ }\href {\doibase
  https://doi.org/10.1103/PhysRevX.2.031001} {\bibfield  {journal} {\bibinfo
  {journal} {Physical Review X}\ }\textbf {\bibinfo {volume} {2}},\ \bibinfo
  {pages} {031001} (\bibinfo {year} {2012})}\BibitemShut {NoStop}%
\bibitem [{\citenamefont {Barson}\ \emph {et~al.}(2019)\citenamefont {Barson},
  \citenamefont {Reddy}, \citenamefont {Yang}, \citenamefont {Manson},
  \citenamefont {Wrachtrup},\ and\ \citenamefont
  {Doherty}}]{barson2019temperature}%
  \BibitemOpen
  \bibfield  {author} {\bibinfo {author} {\bibfnamefont {M.~S.~J.}\
  \bibnamefont {Barson}}, \bibinfo {author} {\bibfnamefont {P.}~\bibnamefont
  {Reddy}}, \bibinfo {author} {\bibfnamefont {S.}~\bibnamefont {Yang}},
  \bibinfo {author} {\bibfnamefont {N.~B.}\ \bibnamefont {Manson}}, \bibinfo
  {author} {\bibfnamefont {J.}~\bibnamefont {Wrachtrup}}, \ and\ \bibinfo
  {author} {\bibfnamefont {M.~W.}\ \bibnamefont {Doherty}},\ }\bibfield
  {title} {\enquote {\bibinfo {title} {Temperature dependence of the c 13
  hyperfine structure of the negatively charged nitrogen-vacancy center in
  diamond},}\ }\href {\doibase https://doi.org/10.1103/PhysRevB.99.094101}
  {\bibfield  {journal} {\bibinfo  {journal} {Physical Review B}\ }\textbf
  {\bibinfo {volume} {99}},\ \bibinfo {pages} {094101} (\bibinfo {year}
  {2019})}\BibitemShut {NoStop}%
\bibitem [{\citenamefont {Plakhotnik}\ and\ \citenamefont
  {Gruber}(2010)}]{plakhotnik2010luminescence}%
  \BibitemOpen
  \bibfield  {author} {\bibinfo {author} {\bibfnamefont {T.}~\bibnamefont
  {Plakhotnik}}\ and\ \bibinfo {author} {\bibfnamefont {D.}~\bibnamefont
  {Gruber}},\ }\bibfield  {title} {\enquote {\bibinfo {title} {Luminescence of
  nitrogen-vacancy centers in nanodiamonds at temperatures between 300 and 700
  k: perspectives on nanothermometry},}\ }\href {\doibase
  https://doi.org/10.1039/C001132K} {\bibfield  {journal} {\bibinfo  {journal}
  {Physical Chemistry Chemical Physics}\ }\textbf {\bibinfo {volume} {12}},\
  \bibinfo {pages} {9751--9756} (\bibinfo {year} {2010})}\BibitemShut {NoStop}%
\bibitem [{\citenamefont {Doherty}\ \emph {et~al.}(2014)\citenamefont
  {Doherty}, \citenamefont {Struzhkin}, \citenamefont {Simpson}, \citenamefont
  {McGuinness}, \citenamefont {Meng}, \citenamefont {Stacey}, \citenamefont
  {Karle}, \citenamefont {Hemley}, \citenamefont {Manson}, \citenamefont
  {Hollenberg},\ and\ \citenamefont {Prawer}}]{doherty2014electronic}%
  \BibitemOpen
  \bibfield  {author} {\bibinfo {author} {\bibfnamefont {M.~W.}\ \bibnamefont
  {Doherty}}, \bibinfo {author} {\bibfnamefont {V.~V.}\ \bibnamefont
  {Struzhkin}}, \bibinfo {author} {\bibfnamefont {D.~A.}\ \bibnamefont
  {Simpson}}, \bibinfo {author} {\bibfnamefont {L.~P.}\ \bibnamefont
  {McGuinness}}, \bibinfo {author} {\bibfnamefont {Y.}~\bibnamefont {Meng}},
  \bibinfo {author} {\bibfnamefont {A.}~\bibnamefont {Stacey}}, \bibinfo
  {author} {\bibfnamefont {T.~J.}\ \bibnamefont {Karle}}, \bibinfo {author}
  {\bibfnamefont {R.~J.}\ \bibnamefont {Hemley}}, \bibinfo {author}
  {\bibfnamefont {N.~B.}\ \bibnamefont {Manson}}, \bibinfo {author}
  {\bibfnamefont {L.~C.~L.}\ \bibnamefont {Hollenberg}}, \ and\ \bibinfo
  {author} {\bibfnamefont {S.}~\bibnamefont {Prawer}},\ }\bibfield  {title}
  {\enquote {\bibinfo {title} {Electronic properties and metrology applications
  of the diamond nv- center under pressure},}\ }\href {\doibase
  https://doi.org/10.1103/PhysRevLett.112.047601} {\bibfield  {journal}
  {\bibinfo  {journal} {Physical review letters}\ }\textbf {\bibinfo {volume}
  {112}},\ \bibinfo {pages} {047601} (\bibinfo {year} {2014})}\BibitemShut
  {NoStop}%
\bibitem [{\citenamefont {Barson}\ \emph {et~al.}(2017)\citenamefont {Barson},
  \citenamefont {Peddibhotla}, \citenamefont {Ovartchaiyapong}, \citenamefont
  {Ganesan}, \citenamefont {Taylor}, \citenamefont {Gebert}, \citenamefont
  {Mielens}, \citenamefont {Koslowski}, \citenamefont {Simpson}, \citenamefont
  {McGuinness}, \citenamefont {McCallum}, \citenamefont {Prawer}, \citenamefont
  {Onoda}, \citenamefont {Ohshima}, \citenamefont {Bleszynski~Jayich},
  \citenamefont {Jelezko}, \citenamefont {Manson}, ,\ and\ \citenamefont
  {Doherty}}]{barson2017nanomechanical}%
  \BibitemOpen
  \bibfield  {author} {\bibinfo {author} {\bibfnamefont {M.~S.~J.}\
  \bibnamefont {Barson}}, \bibinfo {author} {\bibfnamefont {P.}~\bibnamefont
  {Peddibhotla}}, \bibinfo {author} {\bibfnamefont {P.}~\bibnamefont
  {Ovartchaiyapong}}, \bibinfo {author} {\bibfnamefont {K.}~\bibnamefont
  {Ganesan}}, \bibinfo {author} {\bibfnamefont {R.~L.}\ \bibnamefont {Taylor}},
  \bibinfo {author} {\bibfnamefont {M.}~\bibnamefont {Gebert}}, \bibinfo
  {author} {\bibfnamefont {Z.}~\bibnamefont {Mielens}}, \bibinfo {author}
  {\bibfnamefont {B.}~\bibnamefont {Koslowski}}, \bibinfo {author}
  {\bibfnamefont {D.~A.}\ \bibnamefont {Simpson}}, \bibinfo {author}
  {\bibfnamefont {L.~P.}\ \bibnamefont {McGuinness}}, \bibinfo {author}
  {\bibfnamefont {J.}~\bibnamefont {McCallum}}, \bibinfo {author}
  {\bibfnamefont {S.}~\bibnamefont {Prawer}}, \bibinfo {author} {\bibfnamefont
  {S.}~\bibnamefont {Onoda}}, \bibinfo {author} {\bibfnamefont
  {T.}~\bibnamefont {Ohshima}}, \bibinfo {author} {\bibfnamefont {A.~C.}\
  \bibnamefont {Bleszynski~Jayich}}, \bibinfo {author} {\bibfnamefont
  {F.}~\bibnamefont {Jelezko}}, \bibinfo {author} {\bibfnamefont {N.~B.}\
  \bibnamefont {Manson}}, , \ and\ \bibinfo {author} {\bibfnamefont {M.~W.}\
  \bibnamefont {Doherty}},\ }\bibfield  {title} {\enquote {\bibinfo {title}
  {Nanomechanical sensing using spins in diamond},}\ }\href {\doibase
  https://doi.org/10.1021/acs.nanolett.6b04544} {\bibfield  {journal} {\bibinfo
   {journal} {Nano letters}\ }\textbf {\bibinfo {volume} {17}},\ \bibinfo
  {pages} {1496--1503} (\bibinfo {year} {2017})}\BibitemShut {NoStop}%
\bibitem [{\citenamefont {Duan}\ \emph {et~al.}(2019)\citenamefont {Duan},
  \citenamefont {Du}, \citenamefont {Kavatamane}, \citenamefont {Arumugam},
  \citenamefont {Tzeng}, \citenamefont {Chang},\ and\ \citenamefont
  {Balasubramanian}}]{duan2019efficient}%
  \BibitemOpen
  \bibfield  {author} {\bibinfo {author} {\bibfnamefont {D.}~\bibnamefont
  {Duan}}, \bibinfo {author} {\bibfnamefont {G.~X.}\ \bibnamefont {Du}},
  \bibinfo {author} {\bibfnamefont {V.~K.}\ \bibnamefont {Kavatamane}},
  \bibinfo {author} {\bibfnamefont {S.}~\bibnamefont {Arumugam}}, \bibinfo
  {author} {\bibfnamefont {Y.-K.}\ \bibnamefont {Tzeng}}, \bibinfo {author}
  {\bibfnamefont {H.-C.}\ \bibnamefont {Chang}}, \ and\ \bibinfo {author}
  {\bibfnamefont {G.}~\bibnamefont {Balasubramanian}},\ }\bibfield  {title}
  {\enquote {\bibinfo {title} {Efficient nitrogen-vacancy centers fluorescence
  excitation and collection from micrometer-sized diamond by a tapered optical
  fiber in endoscope-type configuration},}\ }\href {\doibase
  https://doi.org/10.1364/OE.27.006734} {\bibfield  {journal} {\bibinfo
  {journal} {Optics Express}\ }\textbf {\bibinfo {volume} {27}},\ \bibinfo
  {pages} {6734--6745} (\bibinfo {year} {2019})}\BibitemShut {NoStop}%
\bibitem [{\citenamefont {Homrighausen}\ \emph {et~al.}(2023)\citenamefont
  {Homrighausen}, \citenamefont {Horsthemke}, \citenamefont {Pogorzelski},
  \citenamefont {Trinschek}, \citenamefont {Gl{\"o}sek{\"o}tter},\ and\
  \citenamefont {Gregor}}]{homrighausen2023edge}%
  \BibitemOpen
  \bibfield  {author} {\bibinfo {author} {\bibfnamefont {J.}~\bibnamefont
  {Homrighausen}}, \bibinfo {author} {\bibfnamefont {L.}~\bibnamefont
  {Horsthemke}}, \bibinfo {author} {\bibfnamefont {J.}~\bibnamefont
  {Pogorzelski}}, \bibinfo {author} {\bibfnamefont {S.}~\bibnamefont
  {Trinschek}}, \bibinfo {author} {\bibfnamefont {P.}~\bibnamefont
  {Gl{\"o}sek{\"o}tter}}, \ and\ \bibinfo {author} {\bibfnamefont
  {M.}~\bibnamefont {Gregor}},\ }\bibfield  {title} {\enquote {\bibinfo {title}
  {Edge-machine-learning-assisted robust magnetometer based on randomly
  oriented nv-ensembles in diamond},}\ }\href {\doibase
  https://doi.org/10.3390/s23031119} {\bibfield  {journal} {\bibinfo  {journal}
  {Sensors}\ }\textbf {\bibinfo {volume} {23}},\ \bibinfo {pages} {1119}
  (\bibinfo {year} {2023})}\BibitemShut {NoStop}%
\bibitem [{\citenamefont {Filipkowski}\ \emph {et~al.}(2022)\citenamefont
  {Filipkowski}, \citenamefont {Mr{\'o}zek}, \citenamefont {St{\k{e}}pniewski},
  \citenamefont {G{\l}owacki}, \citenamefont {Pysz}, \citenamefont {Gawlik},
  \citenamefont {Buczy{\'n}ski}, \citenamefont {Klimczak},\ and\ \citenamefont
  {Wojciechowski}}]{filipkowski2022magnetically}%
  \BibitemOpen
  \bibfield  {author} {\bibinfo {author} {\bibfnamefont {A.}~\bibnamefont
  {Filipkowski}}, \bibinfo {author} {\bibfnamefont {M.}~\bibnamefont
  {Mr{\'o}zek}}, \bibinfo {author} {\bibfnamefont {G.}~\bibnamefont
  {St{\k{e}}pniewski}}, \bibinfo {author} {\bibfnamefont {M.}~\bibnamefont
  {G{\l}owacki}}, \bibinfo {author} {\bibfnamefont {D.}~\bibnamefont {Pysz}},
  \bibinfo {author} {\bibfnamefont {W.}~\bibnamefont {Gawlik}}, \bibinfo
  {author} {\bibfnamefont {R.}~\bibnamefont {Buczy{\'n}ski}}, \bibinfo {author}
  {\bibfnamefont {M.}~\bibnamefont {Klimczak}}, \ and\ \bibinfo {author}
  {\bibfnamefont {A.}~\bibnamefont {Wojciechowski}},\ }\bibfield  {title}
  {\enquote {\bibinfo {title} {Magnetically sensitive fiber probe with
  nitrogen-vacancy center nanodiamonds integrated in a suspended core},}\
  }\href {\doibase https://doi.org/10.1364/OE.458162} {\bibfield  {journal}
  {\bibinfo  {journal} {Optics Express}\ }\textbf {\bibinfo {volume} {30}},\
  \bibinfo {pages} {19573--19581} (\bibinfo {year} {2022})}\BibitemShut
  {NoStop}%
\bibitem [{\citenamefont {Zhang}\ \emph {et~al.}(2021)\citenamefont {Zhang},
  \citenamefont {Dong}, \citenamefont {Du}, \citenamefont {Lin}, \citenamefont
  {Li}, \citenamefont {Zhu}, \citenamefont {Wang}, \citenamefont {Chen},
  \citenamefont {Guo},\ and\ \citenamefont {Sun}}]{zhang2021robust}%
  \BibitemOpen
  \bibfield  {author} {\bibinfo {author} {\bibfnamefont {S.-C.}\ \bibnamefont
  {Zhang}}, \bibinfo {author} {\bibfnamefont {Y.}~\bibnamefont {Dong}},
  \bibinfo {author} {\bibfnamefont {B.}~\bibnamefont {Du}}, \bibinfo {author}
  {\bibfnamefont {H.-B.}\ \bibnamefont {Lin}}, \bibinfo {author} {\bibfnamefont
  {S.}~\bibnamefont {Li}}, \bibinfo {author} {\bibfnamefont {W.}~\bibnamefont
  {Zhu}}, \bibinfo {author} {\bibfnamefont {G.-Z.}\ \bibnamefont {Wang}},
  \bibinfo {author} {\bibfnamefont {X.-D.}\ \bibnamefont {Chen}}, \bibinfo
  {author} {\bibfnamefont {G.-C.}\ \bibnamefont {Guo}}, \ and\ \bibinfo
  {author} {\bibfnamefont {F.-W.}\ \bibnamefont {Sun}},\ }\bibfield  {title}
  {\enquote {\bibinfo {title} {A robust fiber-based quantum thermometer coupled
  with nitrogen-vacancy centers},}\ }\href {\doibase
  https://doi.org/10.1063/5.0044824} {\bibfield  {journal} {\bibinfo  {journal}
  {Review of Scientific Instruments}\ }\textbf {\bibinfo {volume} {92}},\
  \bibinfo {pages} {044904} (\bibinfo {year} {2021})}\BibitemShut {NoStop}%
\bibitem [{\citenamefont {Dix}\ \emph {et~al.}(2024)\citenamefont {Dix},
  \citenamefont {Barbosa}, \citenamefont {Gutsche}, \citenamefont
  {Witzenrath},\ and\ \citenamefont {Widera}}]{dix2024miniaturized}%
  \BibitemOpen
  \bibfield  {author} {\bibinfo {author} {\bibfnamefont {D.}~\bibnamefont
  {Dix}, \bibfnamefont {S.and~L{\"o}nard}}, \bibinfo {author} {\bibfnamefont
  {I.~C.}\ \bibnamefont {Barbosa}}, \bibinfo {author} {\bibfnamefont
  {J.}~\bibnamefont {Gutsche}}, \bibinfo {author} {\bibfnamefont
  {J.}~\bibnamefont {Witzenrath}}, \ and\ \bibinfo {author} {\bibfnamefont
  {A.}~\bibnamefont {Widera}},\ }\bibfield  {title} {\enquote {\bibinfo {title}
  {A miniaturized magnetic field sensor based on nitrogen-vacancy centers},}\
  }\href {https://doi.org/10.48550/arXiv.2402.19372} {\bibfield  {journal}
  {\bibinfo  {journal} {arXiv preprint arXiv:2402.19372}\ } (\bibinfo {year}
  {2024})}\BibitemShut {NoStop}%
\bibitem [{\citenamefont {Zhao}\ \emph {et~al.}(2019)\citenamefont {Zhao},
  \citenamefont {Guo}, \citenamefont {Zhao}, \citenamefont {Du}, \citenamefont
  {Li}, \citenamefont {Wang}, \citenamefont {Wu}, \citenamefont {Chen},
  \citenamefont {Tang},\ and\ \citenamefont {Liu}}]{zhao2019high}%
  \BibitemOpen
  \bibfield  {author} {\bibinfo {author} {\bibfnamefont {B.}~\bibnamefont
  {Zhao}}, \bibinfo {author} {\bibfnamefont {H.}~\bibnamefont {Guo}}, \bibinfo
  {author} {\bibfnamefont {R.}~\bibnamefont {Zhao}}, \bibinfo {author}
  {\bibfnamefont {F.}~\bibnamefont {Du}}, \bibinfo {author} {\bibfnamefont
  {Z.}~\bibnamefont {Li}}, \bibinfo {author} {\bibfnamefont {L.}~\bibnamefont
  {Wang}}, \bibinfo {author} {\bibfnamefont {D.}~\bibnamefont {Wu}}, \bibinfo
  {author} {\bibfnamefont {Y.}~\bibnamefont {Chen}}, \bibinfo {author}
  {\bibfnamefont {J.}~\bibnamefont {Tang}}, \ and\ \bibinfo {author}
  {\bibfnamefont {J.}~\bibnamefont {Liu}},\ }\bibfield  {title} {\enquote
  {\bibinfo {title} {High-sensitivity three-axis vector magnetometry using
  electron spin ensembles in single-crystal diamond},}\ }\href {\doibase
  10.1109/LMAG.2019.2891616} {\bibfield  {journal} {\bibinfo  {journal} {IEEE
  Magnetics Letters}\ }\textbf {\bibinfo {volume} {10}},\ \bibinfo {pages}
  {1--4} (\bibinfo {year} {2019})}\BibitemShut {NoStop}%
\bibitem [{\citenamefont {Dmitriev}\ and\ \citenamefont
  {Vershovskii}(2016)}]{dmitriev2016concept}%
  \BibitemOpen
  \bibfield  {author} {\bibinfo {author} {\bibfnamefont {A.~K.}\ \bibnamefont
  {Dmitriev}}\ and\ \bibinfo {author} {\bibfnamefont {A.~K.}\ \bibnamefont
  {Vershovskii}},\ }\bibfield  {title} {\enquote {\bibinfo {title} {Concept of
  a microscale vector magnetic field sensor based on nitrogen-vacancy centers
  in diamond},}\ }\href {\doibase https://doi.org/10.1364/JOSAB.33.0000B1}
  {\bibfield  {journal} {\bibinfo  {journal} {Journal of the Optical Society of
  America B}\ }\textbf {\bibinfo {volume} {33}},\ \bibinfo {pages} {B1--B4}
  (\bibinfo {year} {2016})}\BibitemShut {NoStop}%
\bibitem [{\citenamefont {Vershovskii}\ and\ \citenamefont
  {Dmitriev}(2015)}]{vershovskii2015micro}%
  \BibitemOpen
  \bibfield  {author} {\bibinfo {author} {\bibfnamefont {A.~K.}\ \bibnamefont
  {Vershovskii}}\ and\ \bibinfo {author} {\bibfnamefont {A.~K.}\ \bibnamefont
  {Dmitriev}},\ }\bibfield  {title} {\enquote {\bibinfo {title} {Micro-scale
  three-component quantum magnetometer based on nitrogen-vacancy color centers
  in diamond crystal},}\ }\href {\doibase
  https://doi.org/10.1134/S1063785015040306} {\bibfield  {journal} {\bibinfo
  {journal} {Technical Physics Letters}\ }\textbf {\bibinfo {volume} {41}},\
  \bibinfo {pages} {393--396} (\bibinfo {year} {2015})}\BibitemShut {NoStop}%
\bibitem [{\citenamefont {Davis}\ \emph {et~al.}(2018)\citenamefont {Davis},
  \citenamefont {Ramesh}, \citenamefont {Bhatnagar}, \citenamefont
  {Lee-Gosselin}, \citenamefont {Barry}, \citenamefont {Glenn}, \citenamefont
  {Walsworth},\ and\ \citenamefont {Shapiro}}]{davis2018mapping}%
  \BibitemOpen
  \bibfield  {author} {\bibinfo {author} {\bibfnamefont {H.~C.}\ \bibnamefont
  {Davis}}, \bibinfo {author} {\bibfnamefont {P.}~\bibnamefont {Ramesh}},
  \bibinfo {author} {\bibfnamefont {A.}~\bibnamefont {Bhatnagar}}, \bibinfo
  {author} {\bibfnamefont {A.}~\bibnamefont {Lee-Gosselin}}, \bibinfo {author}
  {\bibfnamefont {J.~F.}\ \bibnamefont {Barry}}, \bibinfo {author}
  {\bibfnamefont {D.~R.}\ \bibnamefont {Glenn}}, \bibinfo {author}
  {\bibfnamefont {R.~L.}\ \bibnamefont {Walsworth}}, \ and\ \bibinfo {author}
  {\bibfnamefont {M.~G.}\ \bibnamefont {Shapiro}},\ }\bibfield  {title}
  {\enquote {\bibinfo {title} {Mapping the microscale origins of magnetic
  resonance image contrast with subcellular diamond magnetometry},}\ }\href
  {\doibase https://doi.org/10.1038/s41467-017-02471-7} {\bibfield  {journal}
  {\bibinfo  {journal} {Nature communications}\ }\textbf {\bibinfo {volume}
  {9}},\ \bibinfo {pages} {131} (\bibinfo {year} {2018})}\BibitemShut {NoStop}%
\bibitem [{\citenamefont {Wang}\ \emph {et~al.}(2015)\citenamefont {Wang},
  \citenamefont {Yuan}, \citenamefont {Huang}, \citenamefont {Rong},
  \citenamefont {Wang}, \citenamefont {Xu}, \citenamefont {Duan}, \citenamefont
  {Ju}, \citenamefont {Shi},\ and\ \citenamefont {Du}}]{wang2015high}%
  \BibitemOpen
  \bibfield  {author} {\bibinfo {author} {\bibfnamefont {P.}~\bibnamefont
  {Wang}}, \bibinfo {author} {\bibfnamefont {Z.}~\bibnamefont {Yuan}}, \bibinfo
  {author} {\bibfnamefont {P.}~\bibnamefont {Huang}}, \bibinfo {author}
  {\bibfnamefont {X.}~\bibnamefont {Rong}}, \bibinfo {author} {\bibfnamefont
  {M.}~\bibnamefont {Wang}}, \bibinfo {author} {\bibfnamefont {X.}~\bibnamefont
  {Xu}}, \bibinfo {author} {\bibfnamefont {C.}~\bibnamefont {Duan}}, \bibinfo
  {author} {\bibfnamefont {C.}~\bibnamefont {Ju}}, \bibinfo {author}
  {\bibfnamefont {F.}~\bibnamefont {Shi}}, \ and\ \bibinfo {author}
  {\bibfnamefont {J.}~\bibnamefont {Du}},\ }\bibfield  {title} {\enquote
  {\bibinfo {title} {High-resolution vector microwave magnetometry based on
  solid-state spins in diamond},}\ }\href {\doibase
  https://doi.org/10.1038/ncomms7631} {\bibfield  {journal} {\bibinfo
  {journal} {Nature communications}\ }\textbf {\bibinfo {volume} {6}},\
  \bibinfo {pages} {6631} (\bibinfo {year} {2015})}\BibitemShut {NoStop}%
\bibitem [{\citenamefont {Clevenson}\ \emph {et~al.}(2018)\citenamefont
  {Clevenson}, \citenamefont {Pham}, \citenamefont {Teale}, \citenamefont
  {Johnson}, \citenamefont {Englund},\ and\ \citenamefont
  {Braje}}]{clevenson2018robust}%
  \BibitemOpen
  \bibfield  {author} {\bibinfo {author} {\bibfnamefont {H.}~\bibnamefont
  {Clevenson}}, \bibinfo {author} {\bibfnamefont {L.~M.}\ \bibnamefont {Pham}},
  \bibinfo {author} {\bibfnamefont {C.}~\bibnamefont {Teale}}, \bibinfo
  {author} {\bibfnamefont {K.}~\bibnamefont {Johnson}}, \bibinfo {author}
  {\bibfnamefont {D.}~\bibnamefont {Englund}}, \ and\ \bibinfo {author}
  {\bibfnamefont {D.}~\bibnamefont {Braje}},\ }\bibfield  {title} {\enquote
  {\bibinfo {title} {Robust high-dynamic-range vector magnetometry with
  nitrogen-vacancy centers in diamond},}\ }\href {\doibase
  https://doi.org/10.1063/1.5034216} {\bibfield  {journal} {\bibinfo  {journal}
  {Applied Physics Letters}\ }\textbf {\bibinfo {volume} {112}},\ \bibinfo
  {pages} {252406} (\bibinfo {year} {2018})}\BibitemShut {NoStop}%
\bibitem [{\citenamefont {Wickenbrock}\ \emph {et~al.}(2021)\citenamefont
  {Wickenbrock}, \citenamefont {Zheng}, \citenamefont {Chatzidrosos},
  \citenamefont {Rebeirro}, \citenamefont {Schneemann},\ and\ \citenamefont
  {Bluemler}}]{wickenbrock2021high}%
  \BibitemOpen
  \bibfield  {author} {\bibinfo {author} {\bibfnamefont {A.}~\bibnamefont
  {Wickenbrock}}, \bibinfo {author} {\bibfnamefont {H.}~\bibnamefont {Zheng}},
  \bibinfo {author} {\bibfnamefont {G.}~\bibnamefont {Chatzidrosos}}, \bibinfo
  {author} {\bibfnamefont {J.~S.}\ \bibnamefont {Rebeirro}}, \bibinfo {author}
  {\bibfnamefont {T.}~\bibnamefont {Schneemann}}, \ and\ \bibinfo {author}
  {\bibfnamefont {P.}~\bibnamefont {Bluemler}},\ }\bibfield  {title} {\enquote
  {\bibinfo {title} {High homogeneity permanent magnet for diamond
  magnetometry},}\ }\href {\doibase https://doi.org/10.1016/j.jmr.2020.106867}
  {\bibfield  {journal} {\bibinfo  {journal} {Journal of Magnetic Resonance}\
  }\textbf {\bibinfo {volume} {322}},\ \bibinfo {pages} {106867} (\bibinfo
  {year} {2021})}\BibitemShut {NoStop}%
\bibitem [{\citenamefont {Chen}\ \emph {et~al.}(2019)\citenamefont {Chen},
  \citenamefont {He}, \citenamefont {Dong}, \citenamefont {Zhao},\ and\
  \citenamefont {Du}}]{chen2019nitrogen}%
  \BibitemOpen
  \bibfield  {author} {\bibinfo {author} {\bibfnamefont {G.-B.}\ \bibnamefont
  {Chen}}, \bibinfo {author} {\bibfnamefont {W.-H.}\ \bibnamefont {He}},
  \bibinfo {author} {\bibfnamefont {M.-M.}\ \bibnamefont {Dong}}, \bibinfo
  {author} {\bibfnamefont {Y.}~\bibnamefont {Zhao}}, \ and\ \bibinfo {author}
  {\bibfnamefont {G.-X.}\ \bibnamefont {Du}},\ }\bibfield  {title} {\enquote
  {\bibinfo {title} {Nitrogen-vacancy axis orientation measurement in diamond
  micro-crystal for tunable rf vectorial field sensing},}\ }\href {\doibase
  10.1109/JSEN.2019.2953359} {\bibfield  {journal} {\bibinfo  {journal} {IEEE
  Sensors Journal}\ }\textbf {\bibinfo {volume} {20}},\ \bibinfo {pages}
  {2440--2445} (\bibinfo {year} {2019})}\BibitemShut {NoStop}%
\bibitem [{\citenamefont {Fukushige}\ \emph {et~al.}(2020)\citenamefont
  {Fukushige}, \citenamefont {Kawaguchi}, \citenamefont {Shimazaki},
  \citenamefont {Tashima}, \citenamefont {Takashima},\ and\ \citenamefont
  {Takeuchi}}]{fukushige2020identification}%
  \BibitemOpen
  \bibfield  {author} {\bibinfo {author} {\bibfnamefont {K.}~\bibnamefont
  {Fukushige}}, \bibinfo {author} {\bibfnamefont {H.}~\bibnamefont
  {Kawaguchi}}, \bibinfo {author} {\bibfnamefont {K.}~\bibnamefont
  {Shimazaki}}, \bibinfo {author} {\bibfnamefont {T.}~\bibnamefont {Tashima}},
  \bibinfo {author} {\bibfnamefont {H.}~\bibnamefont {Takashima}}, \ and\
  \bibinfo {author} {\bibfnamefont {S.}~\bibnamefont {Takeuchi}},\ }\bibfield
  {title} {\enquote {\bibinfo {title} {Identification of the orientation of a
  single nv center in a nanodiamond using a three-dimensionally controlled
  magnetic field},}\ }\href {\doibase https://doi.org/10.1063/5.0009698}
  {\bibfield  {journal} {\bibinfo  {journal} {Applied Physics Letters}\
  }\textbf {\bibinfo {volume} {116}},\ \bibinfo {pages} {264002} (\bibinfo
  {year} {2020})}\BibitemShut {NoStop}%
\bibitem [{\citenamefont {Li}\ \emph {et~al.}(2023)\citenamefont {Li},
  \citenamefont {Zhang}, \citenamefont {Guo}, \citenamefont {Guo},
  \citenamefont {Yu}, \citenamefont {Zhang}, \citenamefont {Wang},
  \citenamefont {Gao},\ and\ \citenamefont {Zhang}}]{li2023orientation}%
  \BibitemOpen
  \bibfield  {author} {\bibinfo {author} {\bibfnamefont {Z.}~\bibnamefont
  {Li}}, \bibinfo {author} {\bibfnamefont {N.}~\bibnamefont {Zhang}}, \bibinfo
  {author} {\bibfnamefont {J.}~\bibnamefont {Guo}}, \bibinfo {author}
  {\bibfnamefont {Q.}~\bibnamefont {Guo}}, \bibinfo {author} {\bibfnamefont
  {T.}~\bibnamefont {Yu}}, \bibinfo {author} {\bibfnamefont {M.}~\bibnamefont
  {Zhang}}, \bibinfo {author} {\bibfnamefont {G.}~\bibnamefont {Wang}},
  \bibinfo {author} {\bibfnamefont {X.}~\bibnamefont {Gao}}, \ and\ \bibinfo
  {author} {\bibfnamefont {X.}~\bibnamefont {Zhang}},\ }\bibfield  {title}
  {\enquote {\bibinfo {title} {Orientation of the nv centers are determined
  using the cylindrical vector beam array},}\ }\href {\doibase
  https://doi.org/10.1364/OE.483191} {\bibfield  {journal} {\bibinfo  {journal}
  {Optics Express}\ }\textbf {\bibinfo {volume} {31}},\ \bibinfo {pages}
  {9299--9307} (\bibinfo {year} {2023})}\BibitemShut {NoStop}%
\bibitem [{\citenamefont {Wang}\ \emph {et~al.}(2023)\citenamefont {Wang},
  \citenamefont {Zhang}, \citenamefont {Yang}, \citenamefont {Wu},
  \citenamefont {Yu},\ and\ \citenamefont {Chen}}]{wang2023orientation}%
  \BibitemOpen
  \bibfield  {author} {\bibinfo {author} {\bibfnamefont {Y.}~\bibnamefont
  {Wang}}, \bibinfo {author} {\bibfnamefont {R.}~\bibnamefont {Zhang}},
  \bibinfo {author} {\bibfnamefont {Y.}~\bibnamefont {Yang}}, \bibinfo {author}
  {\bibfnamefont {Q.}~\bibnamefont {Wu}}, \bibinfo {author} {\bibfnamefont
  {Z.}~\bibnamefont {Yu}}, \ and\ \bibinfo {author} {\bibfnamefont
  {B.}~\bibnamefont {Chen}},\ }\bibfield  {title} {\enquote {\bibinfo {title}
  {Orientation determination of nitrogen-vacancy center in diamond using a
  static magnetic field},}\ }\href {https://doi.org/10.1088/1674-1056/acc0f7}
  {\bibfield  {journal} {\bibinfo  {journal} {Chinese Physics B}\ } (\bibinfo
  {year} {2023})}\BibitemShut {NoStop}%
\bibitem [{\citenamefont {Blakley}\ \emph {et~al.}(2016)\citenamefont
  {Blakley}, \citenamefont {Fedotov}, \citenamefont {Amitonova}, \citenamefont
  {Serebryannikov}, \citenamefont {Perez}, \citenamefont {Kilin},\ and\
  \citenamefont {Zheltikov}}]{blakley2016fiber}%
  \BibitemOpen
  \bibfield  {author} {\bibinfo {author} {\bibfnamefont {S.~M.}\ \bibnamefont
  {Blakley}}, \bibinfo {author} {\bibfnamefont {I.~V.}\ \bibnamefont
  {Fedotov}}, \bibinfo {author} {\bibfnamefont {L.~V.}\ \bibnamefont
  {Amitonova}}, \bibinfo {author} {\bibfnamefont {E.~E.}\ \bibnamefont
  {Serebryannikov}}, \bibinfo {author} {\bibfnamefont {H.}~\bibnamefont
  {Perez}}, \bibinfo {author} {\bibfnamefont {S.~Y.}\ \bibnamefont {Kilin}}, \
  and\ \bibinfo {author} {\bibfnamefont {A.~M.}\ \bibnamefont {Zheltikov}},\
  }\bibfield  {title} {\enquote {\bibinfo {title} {Fiber-optic vectorial
  magnetic-field gradiometry by a spatiotemporal differential optical detection
  of magnetic resonance in nitrogen--vacancy centers in diamond},}\ }\href
  {\doibase https://doi.org/10.1364/OL.41.002057} {\bibfield  {journal}
  {\bibinfo  {journal} {Optics Letters}\ }\textbf {\bibinfo {volume} {41}},\
  \bibinfo {pages} {2057--2060} (\bibinfo {year} {2016})}\BibitemShut {NoStop}%
\bibitem [{\citenamefont {Fedotov}\ \emph {et~al.}(2014)\citenamefont
  {Fedotov}, \citenamefont {Doronina-Amitonova}, \citenamefont {Voronin},
  \citenamefont {Levchenko}, \citenamefont {Zibrov}, \citenamefont
  {Sidorov-Biryukov}, \citenamefont {Fedotov}, \citenamefont {Velichansky},\
  and\ \citenamefont {Zheltikov}}]{fedotov2014electron}%
  \BibitemOpen
  \bibfield  {author} {\bibinfo {author} {\bibfnamefont {I.~V.}\ \bibnamefont
  {Fedotov}}, \bibinfo {author} {\bibfnamefont {L.~V.}\ \bibnamefont
  {Doronina-Amitonova}}, \bibinfo {author} {\bibfnamefont {A.~A.}\ \bibnamefont
  {Voronin}}, \bibinfo {author} {\bibfnamefont {A.~O.}\ \bibnamefont
  {Levchenko}}, \bibinfo {author} {\bibfnamefont {S.~A.}\ \bibnamefont
  {Zibrov}}, \bibinfo {author} {\bibfnamefont {D.~A.}\ \bibnamefont
  {Sidorov-Biryukov}}, \bibinfo {author} {\bibfnamefont {A.~B.}\ \bibnamefont
  {Fedotov}}, \bibinfo {author} {\bibfnamefont {V.~L.}\ \bibnamefont
  {Velichansky}}, \ and\ \bibinfo {author} {\bibfnamefont {A.~M.}\ \bibnamefont
  {Zheltikov}},\ }\bibfield  {title} {\enquote {\bibinfo {title} {Electron spin
  manipulation and readout through an optical fiber},}\ }\href {\doibase
  https://doi.org/10.1038/srep05362} {\bibfield  {journal} {\bibinfo  {journal}
  {Scientific Reports}\ }\textbf {\bibinfo {volume} {4}},\ \bibinfo {pages}
  {1--6} (\bibinfo {year} {2014})}\BibitemShut {NoStop}%
\bibitem [{\citenamefont {Sasaki}\ \emph {et~al.}(2016)\citenamefont {Sasaki},
  \citenamefont {Monnai}, \citenamefont {Saijo}, \citenamefont {Fujita},
  \citenamefont {Watanabe}, \citenamefont {Ishi-Hayase}, \citenamefont {Itoh},\
  and\ \citenamefont {Abe}}]{Sasaki2016}%
  \BibitemOpen
  \bibfield  {author} {\bibinfo {author} {\bibfnamefont {K.}~\bibnamefont
  {Sasaki}}, \bibinfo {author} {\bibfnamefont {Y.}~\bibnamefont {Monnai}},
  \bibinfo {author} {\bibfnamefont {S.}~\bibnamefont {Saijo}}, \bibinfo
  {author} {\bibfnamefont {R.}~\bibnamefont {Fujita}}, \bibinfo {author}
  {\bibfnamefont {H.}~\bibnamefont {Watanabe}}, \bibinfo {author}
  {\bibfnamefont {J.}~\bibnamefont {Ishi-Hayase}}, \bibinfo {author}
  {\bibfnamefont {K.~M.}\ \bibnamefont {Itoh}}, \ and\ \bibinfo {author}
  {\bibfnamefont {E.}~\bibnamefont {Abe}},\ }\bibfield  {title} {\enquote
  {\bibinfo {title} {Broadband, large-area microwave antenna for optically
  detected magnetic resonance of nitrogen-vacancy centers in diamond},}\ }\href
  {\doibase https://doi.org/10.1063/1.4952418} {\bibfield  {journal} {\bibinfo
  {journal} {Review of Scientific Instruments}\ }\textbf {\bibinfo {volume}
  {87}},\ \bibinfo {pages} {053904} (\bibinfo {year} {2016})}\BibitemShut
  {NoStop}%
\bibitem [{\citenamefont {Chipaux}\ \emph {et~al.}(2015)\citenamefont
  {Chipaux}, \citenamefont {Tallaire}, \citenamefont {Achard}, \citenamefont
  {Pezzagna}, \citenamefont {Meijer}, \citenamefont {Jacques}, \citenamefont
  {Roch},\ and\ \citenamefont {Debuisschert}}]{chipaux2015magnetic}%
  \BibitemOpen
  \bibfield  {author} {\bibinfo {author} {\bibfnamefont {M.}~\bibnamefont
  {Chipaux}}, \bibinfo {author} {\bibfnamefont {A.}~\bibnamefont {Tallaire}},
  \bibinfo {author} {\bibfnamefont {J.}~\bibnamefont {Achard}}, \bibinfo
  {author} {\bibfnamefont {S.}~\bibnamefont {Pezzagna}}, \bibinfo {author}
  {\bibfnamefont {J.}~\bibnamefont {Meijer}}, \bibinfo {author} {\bibfnamefont
  {V.}~\bibnamefont {Jacques}}, \bibinfo {author} {\bibfnamefont {J.-F.}\
  \bibnamefont {Roch}}, \ and\ \bibinfo {author} {\bibfnamefont
  {T.}~\bibnamefont {Debuisschert}},\ }\bibfield  {title} {\enquote {\bibinfo
  {title} {Magnetic imaging with an ensemble of nitrogen-vacancy centers in
  diamond},}\ }\href {\doibase https://doi.org/10.1140/epjd/e2015-60080-1}
  {\bibfield  {journal} {\bibinfo  {journal} {The European Physical Journal D}\
  }\textbf {\bibinfo {volume} {69}},\ \bibinfo {pages} {1--10} (\bibinfo {year}
  {2015})}\BibitemShut {NoStop}%
\bibitem [{\citenamefont {Barry}\ \emph {et~al.}(2020)\citenamefont {Barry},
  \citenamefont {Schloss}, \citenamefont {Bauch}, \citenamefont {Turner},
  \citenamefont {Hart}, \citenamefont {Pham},\ and\ \citenamefont
  {Walsworth}}]{barry2020sensitivity}%
  \BibitemOpen
  \bibfield  {author} {\bibinfo {author} {\bibfnamefont {J.~F.}\ \bibnamefont
  {Barry}}, \bibinfo {author} {\bibfnamefont {J.~M.}\ \bibnamefont {Schloss}},
  \bibinfo {author} {\bibfnamefont {E.}~\bibnamefont {Bauch}}, \bibinfo
  {author} {\bibfnamefont {M.~J.}\ \bibnamefont {Turner}}, \bibinfo {author}
  {\bibfnamefont {C.~A.}\ \bibnamefont {Hart}}, \bibinfo {author}
  {\bibfnamefont {L.~M.}\ \bibnamefont {Pham}}, \ and\ \bibinfo {author}
  {\bibfnamefont {R.~L.}\ \bibnamefont {Walsworth}},\ }\bibfield  {title}
  {\enquote {\bibinfo {title} {Sensitivity optimization for nv-diamond
  magnetometry},}\ }\href {\doibase
  https://doi.org/10.1103/RevModPhys.92.015004} {\bibfield  {journal} {\bibinfo
   {journal} {Reviews of Modern Physics}\ }\textbf {\bibinfo {volume} {92}},\
  \bibinfo {pages} {015004} (\bibinfo {year} {2020})}\BibitemShut {NoStop}%
\bibitem [{\citenamefont {Steinert}\ \emph {et~al.}(2010)\citenamefont
  {Steinert}, \citenamefont {Dolde}, \citenamefont {Neumann}, \citenamefont
  {Aird}, \citenamefont {Naydenov}, \citenamefont {Balasubramanian},
  \citenamefont {Jelezko},\ and\ \citenamefont {Wrachtrup}}]{steinert2010high}%
  \BibitemOpen
  \bibfield  {author} {\bibinfo {author} {\bibfnamefont {S.}~\bibnamefont
  {Steinert}}, \bibinfo {author} {\bibfnamefont {F.}~\bibnamefont {Dolde}},
  \bibinfo {author} {\bibfnamefont {P.}~\bibnamefont {Neumann}}, \bibinfo
  {author} {\bibfnamefont {A.}~\bibnamefont {Aird}}, \bibinfo {author}
  {\bibfnamefont {B.}~\bibnamefont {Naydenov}}, \bibinfo {author}
  {\bibfnamefont {G.}~\bibnamefont {Balasubramanian}}, \bibinfo {author}
  {\bibfnamefont {F.}~\bibnamefont {Jelezko}}, \ and\ \bibinfo {author}
  {\bibfnamefont {J.}~\bibnamefont {Wrachtrup}},\ }\bibfield  {title} {\enquote
  {\bibinfo {title} {High sensitivity magnetic imaging using an array of spins
  in diamond},}\ }\href {\doibase https://doi.org/10.1063/1.3385689} {\bibfield
   {journal} {\bibinfo  {journal} {Review of scientific instruments}\ }\textbf
  {\bibinfo {volume} {81}},\ \bibinfo {pages} {043705} (\bibinfo {year}
  {2010})}\BibitemShut {NoStop}%
\bibitem [{\citenamefont {Nowodzinski}\ \emph {et~al.}(2015)\citenamefont
  {Nowodzinski}, \citenamefont {Chipaux}, \citenamefont {Toraille},
  \citenamefont {Jacques}, \citenamefont {Roch},\ and\ \citenamefont
  {Debuisschert}}]{nowodzinski2015nitrogen}%
  \BibitemOpen
  \bibfield  {author} {\bibinfo {author} {\bibfnamefont {A.}~\bibnamefont
  {Nowodzinski}}, \bibinfo {author} {\bibfnamefont {M.}~\bibnamefont
  {Chipaux}}, \bibinfo {author} {\bibfnamefont {L.}~\bibnamefont {Toraille}},
  \bibinfo {author} {\bibfnamefont {V.}~\bibnamefont {Jacques}}, \bibinfo
  {author} {\bibfnamefont {J.-F.}\ \bibnamefont {Roch}}, \ and\ \bibinfo
  {author} {\bibfnamefont {T.}~\bibnamefont {Debuisschert}},\ }\bibfield
  {title} {\enquote {\bibinfo {title} {Nitrogen-vacancy centers in diamond for
  current imaging at the redistributive layer level of integrated circuits},}\
  }\href {\doibase https://doi.org/10.1016/j.microrel.2015.06.069} {\bibfield
  {journal} {\bibinfo  {journal} {Microelectronics Reliability}\ }\textbf
  {\bibinfo {volume} {55}},\ \bibinfo {pages} {1549--1553} (\bibinfo {year}
  {2015})}\BibitemShut {NoStop}%
\bibitem [{\citenamefont {St{\"u}rner}\ \emph {et~al.}(2021)\citenamefont
  {St{\"u}rner}, \citenamefont {Brenneis}, \citenamefont {Buck}, \citenamefont
  {Kassel}, \citenamefont {R{\"o}lver}, \citenamefont {Fuchs}, \citenamefont
  {Savitsky}, \citenamefont {Suter}, \citenamefont {Grimmel}, \citenamefont
  {Hengesbach}, \citenamefont {F{\"o}rtsch}, \citenamefont {Nakamura},
  \citenamefont {Sumiya}, \citenamefont {Onoda}, \citenamefont {Isoya},\ and\
  \citenamefont {Jelezko}}]{sturner2021integrated}%
  \BibitemOpen
  \bibfield  {author} {\bibinfo {author} {\bibfnamefont {F.~M.}\ \bibnamefont
  {St{\"u}rner}}, \bibinfo {author} {\bibfnamefont {A.}~\bibnamefont
  {Brenneis}}, \bibinfo {author} {\bibfnamefont {T.}~\bibnamefont {Buck}},
  \bibinfo {author} {\bibfnamefont {J.}~\bibnamefont {Kassel}}, \bibinfo
  {author} {\bibfnamefont {R.}~\bibnamefont {R{\"o}lver}}, \bibinfo {author}
  {\bibfnamefont {T.}~\bibnamefont {Fuchs}}, \bibinfo {author} {\bibfnamefont
  {A.}~\bibnamefont {Savitsky}}, \bibinfo {author} {\bibfnamefont
  {D.}~\bibnamefont {Suter}}, \bibinfo {author} {\bibfnamefont
  {J.}~\bibnamefont {Grimmel}}, \bibinfo {author} {\bibfnamefont
  {S.}~\bibnamefont {Hengesbach}}, \bibinfo {author} {\bibfnamefont
  {M.}~\bibnamefont {F{\"o}rtsch}}, \bibinfo {author} {\bibfnamefont
  {K.}~\bibnamefont {Nakamura}}, \bibinfo {author} {\bibfnamefont
  {H.}~\bibnamefont {Sumiya}}, \bibinfo {author} {\bibfnamefont
  {S.}~\bibnamefont {Onoda}}, \bibinfo {author} {\bibfnamefont
  {J.}~\bibnamefont {Isoya}}, \ and\ \bibinfo {author} {\bibfnamefont
  {F.}~\bibnamefont {Jelezko}},\ }\bibfield  {title} {\enquote {\bibinfo
  {title} {Integrated and portable magnetometer based on nitrogen-vacancy
  ensembles in diamond},}\ }\href {\doibase
  https://doi.org/10.1002/qute.202000111} {\bibfield  {journal} {\bibinfo
  {journal} {Advanced Quantum Technologies}\ }\textbf {\bibinfo {volume} {4}},\
  \bibinfo {pages} {2000111} (\bibinfo {year} {2021})}\BibitemShut {NoStop}%
\bibitem [{\citenamefont {Dr{\'e}au}\ \emph {et~al.}(2011)\citenamefont
  {Dr{\'e}au}, \citenamefont {Lesik}, \citenamefont {Rondin}, \citenamefont
  {Spinicelli}, \citenamefont {Arcizet}, \citenamefont {Roch},\ and\
  \citenamefont {Jacques}}]{dreau2011avoiding}%
  \BibitemOpen
  \bibfield  {author} {\bibinfo {author} {\bibfnamefont {A}~\bibnamefont
  {Dr{\'e}au}}, \bibinfo {author} {\bibfnamefont {M}~\bibnamefont {Lesik}},
  \bibinfo {author} {\bibfnamefont {L}~\bibnamefont {Rondin}}, \bibinfo
  {author} {\bibfnamefont {P}~\bibnamefont {Spinicelli}}, \bibinfo {author}
  {\bibfnamefont {O}~\bibnamefont {Arcizet}}, \bibinfo {author} {\bibfnamefont
  {J-F}\ \bibnamefont {Roch}}, \ and\ \bibinfo {author} {\bibfnamefont
  {V}~\bibnamefont {Jacques}},\ }\bibfield  {title} {\enquote {\bibinfo {title}
  {Avoiding power broadening in optically detected magnetic resonance of single
  nv defects for enhanced dc magnetic field sensitivity},}\ }\href {\doibase
  https://doi.org/10.1103/PhysRevB.84.195204} {\bibfield  {journal} {\bibinfo
  {journal} {Physical Review B}\ }\textbf {\bibinfo {volume} {84}},\ \bibinfo
  {pages} {195204} (\bibinfo {year} {2011})}\BibitemShut {NoStop}%
\bibitem [{\citenamefont {Jensen}\ \emph {et~al.}(2013)\citenamefont {Jensen},
  \citenamefont {Acosta}, \citenamefont {Jarmola},\ and\ \citenamefont
  {Budker}}]{jensen2013light}%
  \BibitemOpen
  \bibfield  {author} {\bibinfo {author} {\bibfnamefont {K.}~\bibnamefont
  {Jensen}}, \bibinfo {author} {\bibfnamefont {V.~M.}\ \bibnamefont {Acosta}},
  \bibinfo {author} {\bibfnamefont {A.}~\bibnamefont {Jarmola}}, \ and\
  \bibinfo {author} {\bibfnamefont {D.}~\bibnamefont {Budker}},\ }\bibfield
  {title} {\enquote {\bibinfo {title} {Light narrowing of magnetic resonances
  in ensembles of nitrogen-vacancy centers in diamond},}\ }\href {\doibase
  https://doi.org/10.1103/PhysRevB.87.014115} {\bibfield  {journal} {\bibinfo
  {journal} {Physical review B}\ }\textbf {\bibinfo {volume} {87}},\ \bibinfo
  {pages} {014115} (\bibinfo {year} {2013})}\BibitemShut {NoStop}%
\bibitem [{\citenamefont {Acosta}\ \emph {et~al.}(2010)\citenamefont {Acosta},
  \citenamefont {Bauch}, \citenamefont {Ledbetter}, \citenamefont {Waxman},
  \citenamefont {Bouchard},\ and\ \citenamefont
  {Budker}}]{acosta2010temperature}%
  \BibitemOpen
  \bibfield  {author} {\bibinfo {author} {\bibfnamefont {V.~M.}\ \bibnamefont
  {Acosta}}, \bibinfo {author} {\bibfnamefont {E.}~\bibnamefont {Bauch}},
  \bibinfo {author} {\bibfnamefont {M.~P.}\ \bibnamefont {Ledbetter}}, \bibinfo
  {author} {\bibfnamefont {A.}~\bibnamefont {Waxman}}, \bibinfo {author}
  {\bibfnamefont {L.-S.}\ \bibnamefont {Bouchard}}, \ and\ \bibinfo {author}
  {\bibfnamefont {D.}~\bibnamefont {Budker}},\ }\bibfield  {title} {\enquote
  {\bibinfo {title} {Temperature dependence of the nitrogen-vacancy magnetic
  resonance in diamond},}\ }\href {\doibase
  https://doi.org/10.1103/PhysRevLett.104.070801} {\bibfield  {journal}
  {\bibinfo  {journal} {Physical review letters}\ }\textbf {\bibinfo {volume}
  {104}},\ \bibinfo {pages} {070801} (\bibinfo {year} {2010})}\BibitemShut
  {NoStop}%
\bibitem [{\citenamefont {Doherty}\ \emph {et~al.}(2012)\citenamefont
  {Doherty}, \citenamefont {Dolde}, \citenamefont {Fedder}, \citenamefont
  {Jelezko}, \citenamefont {Wrachtrup}, \citenamefont {Manson},\ and\
  \citenamefont {Hollenberg}}]{doherty2012theory}%
  \BibitemOpen
  \bibfield  {author} {\bibinfo {author} {\bibfnamefont {M.~W.}\ \bibnamefont
  {Doherty}}, \bibinfo {author} {\bibfnamefont {F.}~\bibnamefont {Dolde}},
  \bibinfo {author} {\bibfnamefont {H.}~\bibnamefont {Fedder}}, \bibinfo
  {author} {\bibfnamefont {F.}~\bibnamefont {Jelezko}}, \bibinfo {author}
  {\bibfnamefont {J.}~\bibnamefont {Wrachtrup}}, \bibinfo {author}
  {\bibfnamefont {N.~B.}\ \bibnamefont {Manson}}, \ and\ \bibinfo {author}
  {\bibfnamefont {L.~C.~L.}\ \bibnamefont {Hollenberg}},\ }\bibfield  {title}
  {\enquote {\bibinfo {title} {Theory of the ground-state spin of the nv-
  center in diamond},}\ }\href {\doibase
  https://doi.org/10.1103/PhysRevB.85.205203} {\bibfield  {journal} {\bibinfo
  {journal} {Physical Review B}\ }\textbf {\bibinfo {volume} {85}},\ \bibinfo
  {pages} {205203} (\bibinfo {year} {2012})}\BibitemShut {NoStop}%
\bibitem [{\citenamefont {Dolde}\ \emph {et~al.}(2011)\citenamefont {Dolde},
  \citenamefont {Fedder}, \citenamefont {Doherty}, \citenamefont {N{\"o}bauer},
  \citenamefont {Rempp}, \citenamefont {Balasubramanian}, \citenamefont {Wolf},
  \citenamefont {Reinhard}, \citenamefont {Hollenberg}, \citenamefont
  {Jelezko},\ and\ \citenamefont {Wrachtrup}}]{dolde2011electric}%
  \BibitemOpen
  \bibfield  {author} {\bibinfo {author} {\bibfnamefont {F.}~\bibnamefont
  {Dolde}}, \bibinfo {author} {\bibfnamefont {H.}~\bibnamefont {Fedder}},
  \bibinfo {author} {\bibfnamefont {M.~W.}\ \bibnamefont {Doherty}}, \bibinfo
  {author} {\bibfnamefont {T.}~\bibnamefont {N{\"o}bauer}}, \bibinfo {author}
  {\bibfnamefont {F.}~\bibnamefont {Rempp}}, \bibinfo {author} {\bibfnamefont
  {G.}~\bibnamefont {Balasubramanian}}, \bibinfo {author} {\bibfnamefont
  {T.}~\bibnamefont {Wolf}}, \bibinfo {author} {\bibfnamefont {F.}~\bibnamefont
  {Reinhard}}, \bibinfo {author} {\bibfnamefont {L.~C.~L.}\ \bibnamefont
  {Hollenberg}}, \bibinfo {author} {\bibfnamefont {F.}~\bibnamefont {Jelezko}},
  \ and\ \bibinfo {author} {\bibfnamefont {J.}~\bibnamefont {Wrachtrup}},\
  }\bibfield  {title} {\enquote {\bibinfo {title} {Electric-field sensing using
  single diamond spins},}\ }\href {\doibase https://doi.org/10.1038/nphys1969}
  {\bibfield  {journal} {\bibinfo  {journal} {Nature Physics}\ }\textbf
  {\bibinfo {volume} {7}},\ \bibinfo {pages} {459--463} (\bibinfo {year}
  {2011})}\BibitemShut {NoStop}%
\bibitem [{\citenamefont {Xiang}\ \emph {et~al.}(1997)\citenamefont {Xiang},
  \citenamefont {Sun}, \citenamefont {Fan},\ and\ \citenamefont
  {Gong}}]{xiang1997generalized}%
  \BibitemOpen
  \bibfield  {author} {\bibinfo {author} {\bibfnamefont {Y.}~\bibnamefont
  {Xiang}}, \bibinfo {author} {\bibfnamefont {D.~Y.}\ \bibnamefont {Sun}},
  \bibinfo {author} {\bibfnamefont {W.}~\bibnamefont {Fan}}, \ and\ \bibinfo
  {author} {\bibfnamefont {X.~G.}\ \bibnamefont {Gong}},\ }\bibfield  {title}
  {\enquote {\bibinfo {title} {Generalized simulated annealing algorithm and
  its application to the thomson model},}\ }\href {\doibase
  https://doi.org/10.1016/S0375-9601(97)00474-X} {\bibfield  {journal}
  {\bibinfo  {journal} {Physics Letters A}\ }\textbf {\bibinfo {volume}
  {233}},\ \bibinfo {pages} {216--220} (\bibinfo {year} {1997})}\BibitemShut
  {NoStop}%
\bibitem [{\citenamefont {Dijkstra}(1959)}]{dijkstra1959note}%
  \BibitemOpen
  \bibfield  {author} {\bibinfo {author} {\bibfnamefont {E.~W.}\ \bibnamefont
  {Dijkstra}},\ }\bibfield  {title} {\enquote {\bibinfo {title} {A note on two
  problems in connexion with graphs},}\ }\href {\doibase
  https://doi.org/10.1007/BF01386390} {\bibfield  {journal} {\bibinfo
  {journal} {Numerische Mathematik}\ }\textbf {\bibinfo {volume} {1}},\
  \bibinfo {pages} {269--271} (\bibinfo {year} {1959})}\BibitemShut {NoStop}%
\end{thebibliography}%

\end{document}